\input harvmac.tex

\vskip 1in
\Title{\vbox{\baselineskip12pt
\hbox to \hsize{\hfill}
\hbox to \hsize{\hfill }}}
{\vbox{
	\centerline{\hbox{Higher Spin Contributions
		}}\vskip 5pt
        \centerline{\hbox{ to Holographic Fluid Dynamics in $AdS_5/CFT_4$
		}} } }
\centerline{Seungjin Lee$^{2}$}
\centerline{and}
\centerline{Dimitri Polyakov$^{1,3}$\footnote{$^{\dagger}$}
{polyakov@sogang.ac.kr,  
twistorstring@gmail.com
$^3$ On leave of absence from Poncelet Laboratory, Institute for Information
Transmission Problems (IITP), Bolshoi Karetnyi per., 19/1 Moscow, Russia }}
\medskip
\centerline{\it Center for Quantum Space-Time (CQUeST)$^{1}$}
\centerline{\it and Department of Physics$^{2}$}
\centerline{\it Sogang University}
\centerline{\it Seoul 121-742, Korea}

\vskip .1in

\centerline {\bf Abstract}

We calculate the graviton's $\beta$-function in $AdS$ 
string-theoretic sigma-model,
perturbed by vertex operators for Vasiliev's higher spin gauge fields
in $AdS_5$. The result is given by $\beta_{mn}=R_{mn}+4T_{mn}(g,u)$
(with AdS radius set to 1 and the graviton polarized
along the $AdS_5$ boundary), with the matter stress-energy
tensor given by that of conformal holographic fluid in $d=4$, evaluated
at the temperature given by $T={1\over{\pi}}$. The stress-energy tensor
is given by $T_{mn}={g_{mn}}+4u_mu_n+\sum_{N}T^{(N)}_{mn}$ where $u$ 
is the vector excitation 
satisfying
$u^2=-1$ and $N$ is the order of the gradient expansion in the dissipative
part of the tensor. We calculate the contributions up to $N=2$. 
The higher spin excitations contribute to the $\beta$-function,
ensuring the overall Weyl covariance of the matter stress tensor. 
We conjecture that 
the structure of gradient expansion in $d=4$ conformal
 hydrodynamics at higher orders
 is controlled by the higher spin operator  algebra in $AdS_5$.

\Date{April 2013}

\vfill\eject
\lref\malda{J. Maldacenvva, 
Adv.Theor.Math.Phys. 2 (1998) 231-252}
\lref\wit{E. Witten, 
Adv.Theor.Math.Phys. 2 (1998) 253-291 }
\lref\klebpol{S. Gubser, I. Klebanov, A. M. Polyakov, Phys.Lett. 
B428 (1998) 105-114}
\lref\hvs{M. Vasiliev, arXiv:1203.5554}
\lref\maldaz{J. Maldacena, A. Zhiboedov, arXiv:1112.1016}
\lref\maldazh{J. Maldacena, A. Zhiboedov, arXiv:1204.3882}
\lref\antalf{. de Mello Koch, A. Jevicki, K. Jin, J. P. Rodrigues,
Phys.Rev. D83 (2011) 025006}
\lref\antals{	
R. de Mello Koch, A. Jevicki, K. Jin, J. P. Rodrigues, Q. Ye, arXiv:
1205.4117}
\lref\metsaevf{ R. Metsaev, arXiv: 1112.0976}
\lref\metsaevs{ R. Metsaev, arXiv:1205.3131}
\lref\tsul{P. Dempster, M. Tsulaia, arXiv:1203.5597}
\lref\selfp{S. Lee, D. Polyakov, Phys.Rev. D85 (2012) 106014}
\lref\selfpp{D. Polyakov, Phys.Rev. D84 (2011) 126004} 
\lref\fvf{E.S. Fradkin, M.A. Vasiliev, Nucl. Phys. B 291, 141 (1987)}
\lref\fvs{E.S. Fradkin, M.A. Vasiliev, Phys. Lett. B 189 (1987) 89}
\lref\mmswf{S.W. MacDowell, F. Mansouri, Phys. Rev.Lett. 38 (1977) 739}
\lref\mmsws{K. S. Stelle and P. C. West, Phys. Rev. D 21 (1980) 1466}
\lref\mmswt{C.Preitschopf and M.A.Vasiliev, hep-th/9805127}
\lref\vmaf{M. A. Vasiliev, Sov. J. Nucl. Phys. 32 (1980) 439,
Yad. Fiz. 32 (1980) 855}
\lref\vmas{V. E. Lopatin and M. A. Vasiliev, Mod. Phys. Lett. A 3 (1988) 257}
\lref\vmat{E.S. Fradkin and M.A. Vasiliev, Mod. Phys. Lett. A 3 (1988) 2983}
\lref\vmafth{M. A. Vasiliev, Nucl. Phys. B 616 (2001) 106 }
\lref\bonellio{G. Bonelli, JHEP 0411 (2004) 059}
\lref\hsaone{M. A. Vasiliev, Fortsch. Phys. 36 (1988) 33}
\lref\hsatwo{E.S. Fradkin and M.A. Vasiliev, Mod. Phys. Lett. A 3 (1988) 2983}
\lref\hsathree{S. E. Konstein and M. A. Vasiliev, Nucl. Phys. B 331 (1990) 475}
\lref\hsafour{M.P. Blencowe, Class. Quantum Grav. 6, 443 (1989)}
\lref\hsafive{E. Bergshoeff, M. Blencowe and K. Stelle, 
Comm. Math. Phys. 128 (1990) 213}
\lref\hsasix{E. Sezgin and P. Sundell, Nucl. Phys. B 634 (2002) 120 }
\lref\hsaseven{M. A. Vasiliev, Phys. Rev. D 66 (2002) 066006 }
\lref\soojongf{M. Henneaux, S.-J. Rey, JHEP 1012:007,2010}
\lref\henneaux{J. D. Brown and M. Henneaux, Commun. Math. Phys. 104, 207 (1986)}
\lref\sagnottia{A. Sagnotti, E. Sezgin, P. Sundell, hep-th/0501156}
\lref\fronsdal{C. Fronsdal, Phys. Rev. D18 (1978) 3624}
\lref\coleman{ S. Coleman, J. Mandula, Phys. Rev. 159 (1967) 1251}
\lref\haag{R. Haag, J. Lopuszanski, M. Sohnius, Nucl. Phys B88 (1975)
257}
\lref\weinberg{ S. Weinberg, Phys. Rev. 133(1964) B1049}
\lref\tseytbuch{E. Buchbinder, A. Tseytlin, JHEP 1008:057,2010}
\lref\fradkin{E. Fradkin, M. Vasiliev, Phys. Lett. B189 (1987) 89}
\lref\skvortsov{E. Skvortsov, M. Vasiliev, Nucl.Phys.B756:117-147 (2006)}
\lref\mva{M. Vasiliev, Phys. Lett. B243 (1990) 378}
\lref\mvb{M. Vasiliev, Int. J. Mod. Phys. D5
(1996) 763}
\lref\mvc{M. Vasiliev, Phys. Lett. B567 (2003) 139}
\lref\brink{A. Bengtsson, I. Bengtsson, L. Brink, Nucl. Phys. B227
 (1983) 31}
\lref\deser{S. Deser, Z. Yang, Class. Quant. Grav 7 (1990) 1491}
\lref\bengt{ A. Bengtsson, I. Bengtsson, N. Linden,
Class. Quant. Grav. 4 (1987) 1333}
\lref\boulanger{ X. Bekaert, N. Boulanger, S. Cnockaert,
J. Math. Phys 46 (2005) 012303}
\lref\bbd{F. Berends, G. Burgers, H. Van Dam ,Nucl.Phys. B260 (1985) 295}
\lref\metsaev{ R. Metsaev, arXiv:0712.3526}
\lref\siegel{ W. Siegel, B. Zwiebach, Nucl. Phys. B282 (1987) 125}
\lref\siegelb{W. Siegel, Nucl. Phys. B 263 (1986) 93}
\lref\nicolai{ A. Neveu, H. Nicolai, P. West, Nucl. Phys. B264 (1986) 573}
\lref\damour{T. Damour, S. Deser, Ann. Poincare Phys. Theor. 47 (1987) 277}
\lref\sagnottib{D. Francia, A. Sagnotti, Phys. Lett. B53 (2002) 303}
\lref\sagnottic{D. Francia, A. Sagnotti, Class. Quant. Grav.
 20 (2003) S473}
\lref\sagnottid{D. Francia, J. Mourad, A. Sagnotti, Nucl. Phys. B773
(2007) 203}
\lref\labastidaa{ J. Labastida, Nucl. Phys. B322 (1989)}
\lref\labastidab{ J. Labastida, Phys. Rev. Lett. 58 (1987) 632}
\lref\mvd{L. Brink, R.Metsaev, M. Vasiliev, Nucl. Phys. B 586 (2000) 183}
\lref\klebanov{ I. Klebanov, A. M. Polyakov,
Phys.Lett.B550 (2002) 213-219}
\lref\mve{
X. Bekaert, S. Cnockaert, C. Iazeolla,
M.A. Vasiliev,  IHES-P-04-47, ULB-TH-04-26, ROM2F-04-29, 
FIAN-TD-17-04, Sep 2005 86pp.}
\lref\sagnottie{A. Campoleoni, D. Francia, J. Mourad, A.
 Sagnotti, Nucl. Phys. B815 (2009) 289-367}
\lref\sagnottif{
A. Campoleoni, D. Francia, J. Mourad, A.
 Sagnotti, arXiv:0904.4447}
\lref\selfa{D. Polyakov, Int.J.Mod.Phys.A20:4001-4020,2005}
\lref\selfb{ D. Polyakov, arXiv:0905.4858}
\lref\selfc{D. Polyakov, arXiv:0906.3663, Int.J.Mod.Phys.A24:6177-6195 (2009)}
\lref\selfd{D. Polyakov, Phys.Rev.D65:084041 (2002)}
\lref\spinself{D. Polyakov, Phys.Rev.D82:066005,2010}
\lref\spinselff{D. Polyakov,Phys.Rev.D83:046005,2011}
\lref\mirian{A. Fotopoulos, M. Tsulaia, Phys.Rev.D76:025014,2007}
\lref\extraa{I. Buchbinder, V. Krykhtin,  arXiv:0707.2181}
\lref\extrac{X. Bekaert, I. Buchbinder, A. Pashnev, M. Tsulaia,
Class.Quant.Grav. 21 (2004) S1457-1464}
\lref\extraf{I. Buchbinder, A. Pashnev, M. Tsulaia, 
Phys.Lett.B523:338-346,2001}
\lref\extrag{I. Buchbinder, E. Fradkin, S. Lyakhovich, V. Pershin,
Phys.Lett. B304 (1993) 239-248}
\lref\bonellia{G. Bonelli, Nucl.Phys.B {669} (2003) 159}
\lref\bonellib{G. Bonelli, JHEP 0311 (2003) 028}
\lref\hsself{D.Polyakov, arXiv:1005.5512}
\lref\sundborg{ B. Sundborg, ucl.Phys.Proc.Suppl. 102 (2001)}
\lref\sezgin{E. Sezgin and P. Sundell,
Nucl.Phys.B644:303- 370,2002}
\lref\giombif{S. Giombi, Xi Yin, arXiv:0912.5105}
\lref\giombis{S. Giombi, Xi Yin, arXiv:1004.3736}
\lref\bekaert{X. Bekaert, N. Boulanger, P. Sundell, arXiv:1007.0435}
\lref\taronna{A. Sagnotti, M. Taronna, arXiv:1006.5242, 
Nucl.Phys.B842:299-361,2011}
\lref\fotopoulos{A. Fotopoulos, M. Tsulaia, arXiv:1007.0747}
\lref\fotopouloss{A. Fotopoulos, M. Tsulaia, arXiv:1009.0727}
\lref\taronnao{M. Taronna, arXiv:1005.3061}
\lref\taronnas{A. Sagnotti, M. Taronna, arXiv:1006.5242 ,
Nucl.Phys.B842:299-361,2011}
\lref\campo{A.Campoleoni,S. Fredenhagen,S. Pfenninger, S. Theisen,
arXiv:1008.4744, JHEP 1011 (2010) 007}
\lref\gaber{M. Gaberdiel, T. Hartman, arXiv:1101.2910, JHEP 1105 (2011) 031}
\lref\per{	
N. Boulanger,S. Leclercq, P. Sundell, JHEP 0808(2008) 056 }
\lref\mav{V. E. Lopatin and M. A. Vasiliev, Mod. Phys. Lett. A 3 (1988) 257}
\lref\zinov{Yu. Zinoviev, Nucl. Phys. B 808 (2009)}
\lref\sv{E.D. Skvortsov, M.A. Vasiliev,
Nucl. Phys.B 756 (2006)117}
\lref\mvasiliev{D.S. Ponomarev, M.A. Vasiliev, Nucl.Phys.B839:466-498,2010}
\lref\zhenya{E.D. Skvortsov, Yu.M. Zinoviev, arXiv:1007.4944}
\lref\perf{N. Boulanger, C. Iazeolla, P. Sundell, JHEP 0907 (2009) 013 }
\lref\pers{N. Boulanger, C. Iazeolla, P. Sundell, JHEP 0907 (2009) 014 }
\lref\selft{D. Polyakov,Phys.Rev.D82:066005,2010}
\lref\selftt{D. Polyakov, Int.J.Mod.Phys.A25:4623-4640,2010}
\lref\tseytlin{I. Klebanov, A Tseytlin, Nucl.Phys.B546:155-181,1999}
\lref\rubenf{R. Manvelyan, K. Mkrtchyan, W. Ruehl, Nucl.Phys.B836:204-221,2010}
\lref\robert{
R. De Mello Koch, A. Jevicki, K. Jin, J. A. P. Rodrigues, arXiv:1008.0633}
\lref\bekae{X. Bekaert, S. Cnockaert, C. Iazeolla, M. A. Vasiliev,
hep-th/0503128}
\lref\vcubic{M. Vasiliev, arXiv:1108.5921}
\lref\sagnottinew{A. Sagnotti, arXiv:1112.4285}
\lref\yin{C.-M. Chang, X. Yin, arXiv:1106.2580 }
\lref\boulskv{ N. Boulanger, E. Skvortsov, arXiv:1107.5028,
JHEP 1109 (2011) 063}
\lref\boulskvz{N. Boulanger, E. Skvortsov,Yu. Zinoviev,
arXiv:1107.1872 , J.Phys.A A44 (2011) 415403
}
\lref\selfsigma{D. Polyakov, Phys.Rev. D84 (2011) 126004}
\lref\wittwist{E. Witten, Commun.Math.Phys. 252 (2004) 189-258}
\lref\soojongs{M. Henneaux, G. L. Gomez, J. Park, S.-J. Rey, 
JHEP 1206 (2012) 037}
\lref\joung{E. Joung, L. Lopez, M. Taronna, JHEP 1207 (2012) 041}
\lref\sor{D. Sorokin, AIP
Conf. Proc. 767 (2005) 172 [hep-th/0405069].
}
\lref\bek{ X. Bekaert, N. Boulanger and Per A. Sundell,
 Rev. Mod. Phys. 84 (2012) 987
}
\lref\sez{E. Sezgin and P. Sundell, JHEP 0507 (2005) 044 [hep-
th/0305040]
} 
\lref\mtar{M. Taronna, JHEP 1204 (2012) 029}
\lref\staro{G. Policastro, D.T. Son, A. Starinets,
Phys.Rev.Lett.87 (2001) 081601}
\lref\start{G. Policastro, D. T. Son, A. Starinets, JHEP 09 (2002) 043}
\lref\starth{P. Kovtun,  D. T. Son, A. Starinets, JHEP 10 (2003) 064}
\lref\buchelf{A. Buchel, J. T. Liu, Phys. Rev. Lett. 93 (2004), 090602}
\lref\buchels{A. Buchel, J. T. Liu, A. Starinets, Nucl. Phys. B707 (2005) 56-68}
\lref\israel{W. Israel, J. M. Stewart, Ann. Phys. 118 (1979) 341-372}

\lref\janik{R. Janik, R. Peschanski, Phys. Rev. D73 (2006) 045013}
\lref\mino{ S. Bhattacharyya, V. Hubeny, S. Minwalla, M. Rangamani,
 JHEP 0802 (2008) 045}
\lref\mint{ S. Bhattacharyya, V. Hubeny, 
R. Loganayagam, G. Mandal,S. Minwalla,
T. Morita, 
 M. Rangamani, H. Reall,JHEP 0806 (2008) 055}
\lref\romasch{R. Baier, P. Romatschke, D.T. Son, A. Starinets, M. Stephanov,
JHEP 0804 (2008) 100}
\lref\wadia{ S. Bhattacharyya, S. Minwalla, S. Wadia,
JHEP 0908 (2009) 059}
\lref\rom{P. Romatschke, Class. Quant. Grav. 26 (2009) 224003}
\lref\buchelt{A. Buchel, R. Myers, M. Paulos, A. Sinha, Phys. Lett. B669 
(2008) 364}
\lref\buchelfr{A. Buchel, R. Myers, A. Sinha, JHEP 0903 (2009) 084}
\lref\buchelff{A. Buchel, R. Myers, JHEP 0908 (2009) 016}
\lref\selfswed{D. Polyakov, arXiv:1207.4751, invited contribution
to special volume on Higher Spin Gauge Theories, to  appear in J. Phys. A}
\lref\euo{E. Joung, L. Lopez, M. Taronna, JHEP 1301 (2013) 168}
\lref\eut{E.  Joung, M. Taronna,  Nucl.Phys. B861 (2012) 145}
\lref\mirianf{P. Dempster, M. Tsulaia, Nucl.Phys. B865 (2012) 353}
\centerline{\bf  1. Introduction}

AdS/CFT  correspondence is known to be an efficient tool to 
investigate dynamics of 
strongly coupled
conformal field theories, such as nonlinear fluid dynamics. For example, 
the equations of hydrodynamics 
can be obtained by deforming the solutions
of gravity with negative cosmological constant and requiring that
the deformations asymptotically satisfy the Einstein equations.
The AdS/hydrodynamics correspondence particularly was used
to calculate various transport 
coefficients in holographic fluid leading to 
remarkable predictions such as the ratio of entropy density
to sheer viscosity in conformal fluid 
~{\staro, \start, \starth,  \janik, \mino, \mint, \romasch, \wadia, \buchelf, 
\buchels, \buchelt,
\buchelfr, \buchelff }
The equations of conformal
hydrodynamics can altogether be cast in the form
of the ``conservation law'':

\eqn\lowen{\nabla_m{T^{mn}}=0}
where
\eqn\grav{\eqalign{T^{mn}=\sum_{N=0}^\infty{T^{mn(N)}}}}
where
\eqn\grav{\eqalign{T^{mn(0)}={1\over3}\epsilon(g^{mn}+4u^mu^n)}}
is the ideal fluid part (with $\epsilon\sim{T}^4$ being
the energy density satisfying $\epsilon=3P$ where $P$ is the pressure
and $T$ is the temperature)
\eqn\grav{\eqalign{T^{mn(1)}=-\eta\rho^{mn}-\zeta\Pi^{mn}
{\vec{\nabla}}{\vec{u}}\cr
\Pi^{mn}=\eta^{mn}+u^mu^n\cr
\rho^{mn}=\Pi^{mp}\Pi^{nq}\nabla_{(p}u_{q)}-{2\over3}\Pi^{mn}\Pi^{pq}\nabla_pu_q}}
being the viscous part (with $\eta$ and $\zeta$ being the shear 
and the bulk viscosities
proportional to the third power of the temperature)
and terms with $N\geq{2}$ representing the dissipative corrections
to the Navier-Stokes equation , 
traceless and transverse, satisfying $T^{mn(N)}u_m=0$, 
which are of order $N$ in the derivatives of $u$ and
 become significant
if the mean free path is comparable to the 
characteristic wavelength in the fluid.

Thus the full stress-energy tensor  in hydrodynamics 
involves the derivative (gradient) expansion
in the velocity with each expansion order producing new transport coefficients.
For example, the second order terms result in 5 new transport coefficients
in conformal hydrodynamics.
At present, there exist various approaches to generate 
the derivative expansion (4) in the dual gravity theories. 
Strictly
speaking, none of these approaches has complete control 
over the expansion (4) and the 
systematic calculation of the relative transport 
coefficients from dual gravity models
is still problematic, especially beyond the second order hydrodynamics 
~{\israel, \rom}
Many gravity models describing the holographic fluids generally involve
 the Gauss-Bonnet terms that
are of higher order in the curvature and the resulting transport 
coefficients particularly depend 
on the Gauss-Bonnet coupling. These theories typically 
have issues with  unitarity and
causality which signals that, in general, they may not
be fundamental but rather  effective theories, 
with certain physical degrees of freedom, such as
higher spins, integrated out. For this reason, 
string theory (which naturally includes higher spin modes)
appears to be a particularly
 promising framework to approach the AdS/hydrodynamics 
duality and to test the transport coefficients
at higher orders. In this paper we analyze the problem of
 AdS/hydrodynamics correspondence from 
 string theory side, by computing graviton's conformal $\beta$-function in  
sigma-model for $AdS_5$ 
noncritical string theory, with the graviton polarized along $d=4$ AdS boundary.
The string model that we use is the RNS string theory perturbed by 
vertex operators describing gravitational perturbations around
$AdS_5$ background
and higher spin gauge fields in Vasiliev's frame-like formalism.  
The  low-energy
limit of this model is given by  the MMSW 
(Mac Dowell - Mansouri - Stelle- West) 
~{\mmswf, \mmsws, \mmswt}
coupled to Vasiliev's higher spin gauge fields 
~{\vmaf, \vmas, \vmat, \vmafth} and the vacuum solution 
of the low-energy theory is given by the AdS geometry ~{\selfsigma}. 
Our main result 
(checked up to $N=2$ level, with higher order checks now being in progress)
is that the beta-function
of the graviton is given by 
\eqn\grav{\eqalign{
\beta_{mn}=R_{mn}+4T_{mn}  \cr
T_{mn}={g_{mn}}+4u_mu_n+\sum_{N=1}^\infty T^{(N)}_{mn}
}}
where $T_{mn}^{(N)}$  are the terms in the 
derivative expansion of the stress-energy
tensor in  $d=4$ hydrodynamics. 
In other words, the low-energy equations in $AdS$ string model
are given by the Einstein equations with cosmological term and the matter, 
with the latter
described by the hydrodynamical stress-energy tensor.
Here $g_{mn}$ and $u_m$ are the massless excitations described 
by spin $2$ and $1$ vertex operators in AdS 
string model, in closed and open string sectors accordingly.
The spin $1$ operators (related to transvection isometry 
generators in AdS space 
~{\selfsigma})
serve as sources of the velocity vector in this model.
As for the open string vertex operators for the massless higher spins, 
in this paper, 
instead of coupling them
to generic Vasiliev's higher spin gauge fields,
we consider the special case of
 coupling these operators to  polynomial
combinations of $u_m$ , constructed to satisfy the
 same linearized on-shell (BRST-invariance) conditions
as the  underlying higher spins. 
As a result, in the leading order of $\alpha^\prime$, the 
structure of the higher order corrections to $\beta_{mn}$ (polynomial in $u$)
 is determined by the structure
constants of the operator algebra of the higher spin vertex operators
(this operator algebra, in turn, fully
controls the cubic couplings for generic higher spins).
In the leading $\alpha^\prime$ order, only the three-point correlation functions 
on the worldsheet contribute to the graviton's $\beta$-function.
Our main result is that, the matter stress tensor 
appearing in the $\beta$-function,
reproduces the derivative expansion (4) 
in the stress tensor of the conformal fluid
at the temperature $T={1\over{\pi}}$, which is checked up to the order of $N=2$.
Since the temperature transforms covariantly under Weyl rescalings, 
this result implies
that the $AdS$ string theory computation reproduces the 
stress tensor of the conformal
fluid at a particular temperature gauge.
We find that, at the order of $N=2$  and higher, the $\beta$-function
receives  nontrivial contributions from higher spin vertex operators.
These contributions are crucial to ensure the conformal covariance of 
the stress tensor.
In particular, at the $N=2$ level the graviton's $\beta$-function
is contributed by the $<2-3-3>$ correlator on the disc, 
while at higher orders operators
of spin 4 and higher also enter the game, so the holographic derivative 
expansion
(4) is controlled by operator algebra of higher spin vertices in the limit
of $\alpha^\prime\rightarrow{0}$.
The rest of the paper is organized as follows:
In the Section 2, we explain the basic vertex operator setup of the
 sigma-model, 
which low-energy limit
describes  the $AdS$ gravity coupled to higher spins in the
 frame-like formalism.
In the section 3, we perform the computations of the $<1-1-2>$ and $<1-3-3>$
correlators, contributing to the graviton's $\beta$-function and reproducing
the holographic expansion (4) up to the second order.
In the concluding section, we comment on the structure of the higher order terms
related to higher spin contributions and 
 and discuss physical
implication of our results.

\centerline{\bf AdS String $\sigma$-Model: Vertex Operators and $2d$ 
Weyl Invariance}

In this section we review the construction of the string-theoretic sigma-model
~{\selfsigma} with some modifications, that will be used in our
calculation of the graviton's beta function.
Technically, the sigma-model that we use in calculations in this work is
similar but not identical the one constructed in our previous works 
(e.g. see ~{\selfsigma}) as it will combine vertex operators
for both Fronsdal-like objects (such as vertex operator 
 for  a graviton describing perturbations around AdS vacuum) and  
those of Vasiliev's type (describing frame-like higher spin excitations
around AdS vacuum solution of the low-energy equations of motion)

The $AdS$ string sigma-model considered in ~{\selfsigma}
was based purely on vertex operators  for frame-like 
gauge fields (rather than those of Fronsdal type) and was  
described by the the generating functional
\eqn\grav{\eqalign{
Z(e_m^a,\omega_m^{ab},\Omega_m^{A_1...A_{s-1}|B_1...B_t})=
\int{D}\lbrack{X,\psi,\varphi,\lambda,ghosts}\rbrack
exp{\lbrace}{-S_{RNS}}+e_m^aF^m{\bar{L}}_a
\cr
+\omega^{ab}_m(p)
(F^m_{b}{\bar{L}}_a-{1\over2}F_{ab}{\bar{L}}^m)+c.c.
\cr
+
\sum_{s\geq{3};0\leq{t}\leq{s-1}} 
\Omega_m^{A_1...A_{s-1}|B_1...B_t}V^m_{A_1...A_{s-1}|B_1...B_t}
\rbrace}}
where
\eqn\grav{\eqalign{S_{RNS}=S_{matter}+S_{bc}+S_{\beta\gamma}+S_{Liouville}\cr
S_{matter}=-{1\over{4\pi}}\int{d^2z}(\partial{X_m}\bar\partial{X^m}
+\psi_m\bar\partial\psi^m+{\bar\psi}_m\partial{\bar\psi}^m)\cr
S_{bc}={1\over{2\pi}}\int{d^2z}(b\bar\partial{c}+{\bar{b}}\partial
{\bar{c}})\cr
S_{\beta\gamma}={1\over{2\pi}}\int{d^2z}(\beta\bar\partial\gamma
+\bar\beta\partial{\bar\gamma})
\cr
S_{Liouville}=-{1\over{4\pi}}\int{d^2z}(\partial\varphi\bar\partial\varphi
+\bar\partial\lambda\lambda+\partial\bar\lambda\bar\lambda
+\mu_0{e^{B\varphi}}(\lambda\bar\lambda+F))
}}
where 
$S_{RNS}$ is the full $d$-dimensional 
RNS superstring action; $X^m(m=0,...{d-1})$ are the space-time coordinates;
$\varphi,\lambda, F$ are components of super Liouville field
and the Liouville background charge is
\eqn\lowen{
Q=B+B^{-1}={\sqrt{{{9-d}\over2}}}}
Next, $e_m^a$ and $\omega_m^{ab}$ are vielbein and spin connection gauge fields
generated by closed string vertex operators which holomorphic and 
antiholomorphic components
are given by
\eqn\grav{\eqalign{
F_m=-2K_{U_1}\circ\int{dz}\lambda\psi_me^{ipX}(z)\cr
U_1=\lambda\psi_me^{ipX}+{i\over2}\gamma\lambda
(({\vec{p}}{\vec{\psi}})\psi_m-p_mP^{(1)}_{\phi-\chi})e^{ipX}}}
or manifestly
\eqn\grav{\eqalign{
F_m=-2\int{dz}{\lbrace}\lambda\psi_m(1-4\partial{c}ce^{2\chi-2\phi})+
\cr
2ce^{\chi-\phi}
(\lambda\partial{X}_m-\partial\varphi\psi_m+q\psi_mP^{(1)}_{\phi-\chi}
-{i\over2}(({\vec{p}}{\vec{\psi}})\psi_m-p_mP^{(1)}_{\phi-\chi}))
\rbrace{e^{ipX}}(z)}}
Next,
\eqn\grav{\eqalign{{\bar{L}}^a=\int{d{\bar{z}}}e^{-3{\bar\phi}}
\lbrace{\bar\lambda}\bar\partial^2{X^a}-2\bar\partial\bar\lambda\bar\partial
{X^a}
\cr+
ip^a({1\over2}\bar\partial^2\bar\lambda+{1\over{q}}
\bar\partial\bar\varphi\bar\partial\bar\lambda
-{1\over2}\bar\lambda(\bar\partial\bar\varphi)^2+
(1+3q^2)\bar\lambda(3\bar\partial\bar\psi_b\bar\psi^b-{1\over{2q}}
\bar\partial^2\bar\varphi))\rbrace{e^{ipX}}}}
at the minimal negative picture $-3$ representation 

and 

\eqn\grav{\eqalign{{\bar{L}}^a=K\circ\int{d{\bar{z}}}e^{{\bar\phi}}
\lbrace{\bar\lambda}\bar\partial^2{X^a}-2\bar\partial\bar\lambda
\bar\partial
{X^a}
\cr+
ip^a({1\over2}\bar\partial^2\bar\lambda+{1\over{q}}
\bar\partial\bar\varphi\bar\partial\bar\lambda
-{1\over2}\bar\lambda(\bar\partial\bar\varphi)^2+
(1+3q^2)\bar\lambda(3\bar\partial\bar\psi_b\bar\psi^b-{1\over{2q}}
\bar\partial^2\bar\varphi))\rbrace{e^{ipX}}}}
at the minimal positive picture $+1$ representation.
(similarly for its holomorphic counterpart $L^a$).
Here and elsewhere below the normalizations of
vertex operators are chosen so as they lead to standard normalizations
of corresponding kinetic terms in low-energy effective action.
Then,
\eqn\grav{\eqalign{
F_{ma}=F_{ma}^{(1)}+F_{ma}^{(2)}+F_{ma}^{(3)}}}
where
\eqn\grav{\eqalign{
F_{ma}^{(1)}=-4qK_{U_2}\circ\int{dz}ce^{\chi-\phi}\lambda\psi_m\psi_a\cr
U_2=\lbrack{Q-Q_3},ce^{\chi-\phi}\lambda\psi_m\psi_a{e^{ipX}}\rbrack
-{i\over2}c\lambda(({\vec{p}}{\vec{\psi}})\psi_a\psi_m-p_m\psi_aP^{(1)}_{\phi-\chi})
e^{ipX}(z)}}
\eqn\grav{\eqalign{
F_{ma}^{(2)}=K\circ\int{dz}\psi_m\psi_a{e^{ipX}}=-4\lbrace{Q},\int{dz}
ce^{2\chi-2\phi}{e^{ipX}}\psi_m\psi_a(z)\rbrace}}
and
\eqn\grav{\eqalign{
F_{ma}^{(3)}=K\circ\int{dz}e^{\phi}(\psi_{\lbrack{m}}\partial^2{X}_{a\rbrack}
-2\partial\psi_{\lbrack{m}}\partial{X}_{a\rbrack})e^{ipX}(z)}}

Here the homotopy transform of an operator $V$
$K\circ{V}$ is defined according to
\eqn\grav{\eqalign{
K{\circ}V=T+{{(-1)^N}\over{N!}}
\oint{{dz}\over{2i\pi}}(z-w)^N:K\partial^N{W}:(z)
\cr
+{1\over{{N!}}}\oint{{dz}\over{2i\pi}}\partial_z^{N+1}{\lbrack}
(z-w)^N{K}(z)\rbrack{K}\lbrace{Q_{brst}},U\rbrace}}
where $w$ is some arbitrary point on the worldsheet,
$U$ and $W$ are the operators defined according 
to
\eqn\grav{\eqalign{\lbrack{Q_{brst}},V(z)\rbrack=\partial{U}(z)+W(z),}}
\eqn\lowen{K=ce^{2\chi-2\phi}}
is the homotopy operator satisfying
${\lbrace}Q_{brst},K{\rbrace}=1$
and $N$ is the leading order of the operator product
\eqn\lowen{K(z_1)W(z_2)\sim{(z_1-z_2)^N}Y(z_2)+O((z_1-z_2)^{N+1})}
The $partial$ homotopy transform 
$T{\rightarrow}L=K_{\Upsilon}\circ{T}$ of an operator $T$ based on
$\Upsilon$ is defined according to 
\eqn\grav{\eqalign{
{{L}}(w)=K_\Upsilon{\circ}T=T+{{(-1)^N}\over{N!}}
\oint{{dz}\over{2i\pi}}(z-w)^N:K\partial^N{\Upsilon}:(z)
\cr
+{1\over{{N!}}}\oint{{dz}\over{2i\pi}}\partial_z^{N+1}{\lbrack}
(z-w)^N{K}(z)\rbrack{K}\lbrace{Q_{brst}},U\rbrace}}
where $N$ is the leading order of the OPE of $K$ and $\Upsilon$. 
Particularly, if $\lbrack{Q_{brst}},T\rbrack=\oint\Upsilon$, 
the partial homotopy transform  obviously
coincides with the usual
homotopy transform.
Finally, $V^m_{a_1...a_{s-1}|b_1...b_t};0\leq{t}\leq{s-1}$ are the 
open string vertex operators for emission of gauge fields
of spin $s$ which, in Vasiliev's approach are described (for each s)
by  collection of two-row fields 
$\Omega^{s-1|t}\equiv\Omega_m^{a_1...a_{s-1}|b_1...b_t}$. 
In this approach, only the $\Omega^{s-1|0}$ field is dynamical 
while those with nonzero $t$ values can
be expressed in terms of order $t$ derivatives of the dynamical field: 
$\Omega^{s-1|t}\sim\partial^{(t)}\Omega^{s-1|0}$
through generalized zero torsion constraints 
(e.g. see ~{\vmaf, \vmas, \vmat, \vmafth, \vcubic})
 In string theory, these constraints are realized in terms of 
ghost cohomology conditions on the higher spin vertex operators 
~{\selfp}.
The BRST-invariance constraints on the hiugher spin  vertex operators (6) 
lead to linearized on-shell constraints on the frame-like fields
while BRST nontriviality conditions lead to gauge symmetry
 transformation by these 
fields; the worldsheet
correlators of the appropriate vertex operators multiplied by the corresponding 
space-time fields are then invariant
by construction ~{\spinselff}.
The vertex operators in the generating functional (6) 
can be classified in terms oh ghost cohomologies
$H_n\sim{H_{-n-2}};n\geq{0}$. For example, the spin 2 operators  
for vielbein and 
connection
gauge fields are the elements of $H_0\otimes{{\bar{H}}_1}+c.c.$
(with $H$ and ${\bar{H}}$) referring to holomorphic and antiholomorphic parts)
while the class of higher spin operators $V_s$ of $s\geq{3}$ that we are 
considering,
 is restricted to
open string vertex operators at nonzero cohomologies; typically, 
$V_s\in{H_n}$ with
$s-2\geq{n}\geq{2s-2}$ (this includes both dynamical and the extra fields
that sit at different cohomologies, with the dynamical field occupying
the lowest order positive cohomology).
In the previous works ~{\selfp, \selfswed}
 we analyzed the low-energy limit of the 
model (6) showing that , in the leading order in $e$ and $\omega$ in the 
absence of
 the open string excitations
(spin 1 and higher spins) its low-energy equations of motion are given by
\eqn\lowen{d\omega+\omega\wedge\omega-{e}\wedge{e}=0}
which vacuum solution is given by $AdS$ geometry (here and elsewhere, unless
specified otherwise we set the AdS radius $\rho_{AdS}=1$).

 All the vertex operators (6) are related to underlying
global symmetries of space-time. In particular, at the limit of momentum zero,
the $L^{a}$-operators, entering expressions for vertex operators of vielbein
 are related to transvection generators in the isometry algebra of $AdS_d$ while
$F^{mn}$-operators are related to the rotational part of this isometry algebra.
 $V^m_{a_1...a_{s-1}|b_1...b_t}$-operators, in turn, are related to the 
higher spin currents, or the generators of the higher
 spin symmetry algebra which, 
to put it roughly,
is the infinite dimensional algebra related to the universal envelopping of the
 isometry algebra.
The higher spin algebra is thus realized in superstring theory as the operator 
algebra of the appropriate
higher spin states, which structure constants are given by the relevant 
3-point correlators, 
or the leading
order contributions to the conformal beta-functions.
 In the present paper, we investigate the correlators
contributing to the $\beta$-function of the graviton. Our interpretation of the 
spin one and the higher spin vertex operators is, however,
different from the one of the previous papers ~{\selfsigma, \selfp}. 
Instead of interpreting the vertex operators  as the emission 
vertices for   fundamental particles, we consider them as sources 
of various polynomials in the vector field $u^m$
(with the polynomial degree obviously related to the spin value) 
with the structure of the polynomials determined
by the on-shell conditions on the corresponding operators. 
The idea is that, in the limit of $\alpha^\prime\rightarrow{0}$
the polynomial contributions to the $\beta$-functions
and the derivative expansion (2) are
controlled by the appropriate  structure constants
in  operator algebra of the higher spin vertex operators
for frame-like gauge fields (which, in turn, naturally  realise
higher spin algebra in a certain basis).

 The constraint $u^2=-1$ 
particularly follows from the on-shell conditions,
allowing us to interpret $u^m$ as the velocity vector in some underlying fluid. 
Then the $\beta$-function equations
of the graviton are realized as the Einstein equations with the cosmological 
term and with the matter, with
the matter stress tensor being that of the hydrodynamics. 
Our claim is that the derivative expansion in holographic $d=4$ hydrodynamics
is determined, in the leading order,
 by the higher spin algebra in $AdS_5$ (calculated in string theory approach), 
with the higher order dissipative terms controlled by the derivative
structure of  higher spin correlators.

\centerline{\bf 3. Graviton in in the Frame-like Sigma-Model}

\centerline{\bf and  2d Weyl
 Invariance of the Operators}

As was explained above, the first building block
that we shall need in our construction is the graviton vertex operator
describing metric perturbations around the $AdS$ vacuum,
as opposed to  operators for vielbeins and spin connections
present in (6)
Similarly to the flat space case (where the graviton operator
is an object bilinear in flat space translation operators), 
the vertex operator for the graviton that  
we are looking for has to be an object
bilinear in $AdS_5$  isometry generators (transvections), 
with the BRST constraints 
imposing appropriate
on-shell conditions and gauge transformations. 
According to (6) there are two types of such operators
- those of $L$-type and those of $F$-type. 
The bilinears of mixed $L-F$ type correspond to
vielbeins and connection gauge fields (elements of 
${\lbrack}H_0\otimes{\bar{H_1}}\rbrack$-cohomology, so the suitable
candidates are either $F-F$ type (in ${\lbrack}H_0\otimes{\bar{H}}_0\rbrack$ 
cohomology)
or $L-L$ type (elements of ${\lbrack}H_1\otimes{\bar{H}}_1\rbrack$). 
The objects of $F-F$ type, however, clearly do not
reproduce proper on-shell conditions and have excessive gauge symmetry, 
therefore the appropriate
candidate for the  graviton operator is the one in
 ${\lbrack}H_1\otimes{\bar{H}}_1\rbrack{\sim}
{\lbrack}H_{-3}\otimes{\bar{H}}_{-3}\rbrack$, with the 
explicit expression given by
\eqn\grav{\eqalign{V_{grav}^{H_{-3}\otimes{H_{-3}}}=G_{mn}(p)c{\bar{c}}e^{-3\phi-3{\bar\phi}}
\lbrace{\bar\lambda}\bar\partial^2{X^m}-2\bar\partial\bar\lambda
\bar\partial
{X^m}
\cr+
ip^m({1\over2}\bar\partial^2\bar\lambda+{1\over{q}}
\partial\varphi\partial\lambda
-{1\over2}\lambda(\partial\varphi)^2+
(1+3q^2)\lambda(3\partial\psi_p\psi^p-{1\over{2q}}
\partial^2\varphi))\rbrace
\lbrace{\bar\lambda}\bar\partial^2{X^n}-2\bar\partial\bar\lambda
\bar\partial
{X^n}
\cr+
ip^n({1\over2}\bar\partial^2\bar\lambda+{1\over{q}}
\bar\partial\bar\varphi\bar\partial\bar\lambda
-{1\over2}\bar\lambda(\bar\partial\bar\varphi)^2+
(1+3q^2)\bar\lambda(3\bar\partial\bar\psi_q\bar\psi^q-{1\over{2q}}
\bar\partial^2\bar\varphi))\rbrace
e^{ipX}}}
at minimal negative picture $-3$ unintegrated  representation
and
\eqn\grav{\eqalign{V=G_{mn}(p){K{\bar{K}}}\circ\int{d^2z}e^{\phi+{\bar\phi}}
\lbrace{\bar\lambda}\bar\partial^2{X^m}-2\bar\partial\bar\lambda
\bar\partial
{X^m}
y\cr+
ip^m({1\over2}\bar\partial^2\bar\lambda+{1\over{q}}
\partial\varphi\partial\lambda
-{1\over2}\lambda(\partial\varphi)^2+
(1+3q^2)\lambda(3\partial\psi_p\psi^p-{1\over{2q}}
\partial^2\varphi))\rbrace
\lbrace{\bar\lambda}\bar\partial^2{X^n}-2\bar\partial\bar\lambda
\bar\partial
{X^n}
\cr+
ip^n({1\over2}\bar\partial^2\bar\lambda+{1\over{q}}
\bar\partial\bar\varphi\bar\partial\bar\lambda
-{1\over2}\bar\lambda(\bar\partial\bar\varphi)^2+
(1+3q^2)\bar\lambda(3\bar\partial\bar\psi_q\bar\psi^q-{1\over{2q}}
\bar\partial^2\bar\varphi))\rbrace
e^{ipX}}}
 at minimal positive picture $+1$ representation (note that the 
operators at positive
cohomologies are always integrated). The antiholomorphic 
${\bar{K}}$-transformation is defined
similarly to the holomorphic one (17).
The transformation
$G^{mn}\rightarrow{G^{mn}}+p^{(m}\Lambda^{n)}$ 
shifts (24) by BRST-exact part.
The leading order contribution to the graviton's beta function is 
the result the Weyl 
invariance constraints on the operator (24). These constraints can be 
conveniently deduced from the OPE:
\eqn\lowen{\sim\int{d^2z}\int{d^2w}T_{z{\bar{z}}}(z,{\bar{z}})V_{grav}(w,{\bar{w}})}
by expanding around the midpoint and  evaluating the coefficient in front of 

$\sim{{{V_{grav}}({{{z+w}\over2}},
{{\bar{z}+{\bar{w}}}\over2})}\over{|z-w|^2}}$
(note that the trace $T_{z{\bar{z}}}$ of the stress-energy tensor, 
generating the Weyl transformation,
is nonzero off-shell or, equivalently, in the underlying $\epsilon$-expansion).
For a usual graviton operator 
$\sim{G_{mn}}(p)\int{d^2w}\partial{X^m}\bar\partial{X^n}e^{ipX}(w,{\bar{w}})$
in the bosonic string this procedure leads, after simple calculation,
 to the standard $\beta$-function contribution, quadratic in momentum,
given by the linearized part of the Ricci tensor plus 
the second derivative of the dilaton
 ${\sim}R_{mn}^{lin.}-2p_mp_n\Phi$ with $\Phi\sim{tr}(G_{mn})$.
The calculation, leading to the identical result, is similar in superstring 
theory.
The graviton operator should then be taken at canonical ghost picture 
(unintegrated $b-c$ picture and
$(-1,-1)$ $\beta-\gamma$ ghost picture), so
$V_{grav}=c{\bar{c}}e^{-\phi-\bar\phi}\psi^m\psi^ne^{ipX}$ and 
the relevant terms in the stress tensor are
\eqn\grav{\eqalign{
T_{z{\bar{z}}}{\equiv}T_{z{\bar{z}}}^{matter}+T_{z{\bar{z}}}^{b-c}+T_{z{\bar{z}}}^{\beta-\gamma}
\cr
=
{1\over2}(-\partial{X_m}\bar\partial{X^m}
-\bar\partial\psi_m\psi^m-\partial\bar\psi_m\bar\psi^m
+\partial\sigma\bar\partial\sigma
+\partial\chi\bar\partial\chi-\partial\phi\bar\partial\phi)}}
The OPE of $V_{grav}$ with $T_{z{\bar{z}}}^{matter}$ then contributes 
the term $\sim{p^2}G_{mn}$ to the graviton's beta-function 
(which is the gauge-fixed
linearized part of the Ricci tensor, with the gauge condition 
$\sim{p^m}G_{mn}=0$), while
the contribution stemming from the OPE with $T_{z{\bar{z}}}^{b-c}$ cancels the one 
from the OPE with
$T_{z{\bar{z}}}^{\beta-\gamma}$ since 
$\partial\sigma\bar\partial\sigma(z,{\bar{z}})c{\bar{c}}(w,{\bar{w}})\sim
{1\over{|z-w|^{2}}}c{\bar{c}}(w,{\bar{w}})$,
$\partial\phi\bar\partial\phi(z,{\bar{z}})e^{-\phi-\bar\phi}(w,{\bar{w}})\sim
{1\over{|z-w|^{2}}}e^{-\phi-\bar\phi}(w,{\bar{w}})$
 and $\sigma$ and $\phi$-terms of
$T_{z{\bar{z}}}$ have opposite signs. It is this cancellation that ensures the 
absence
of ``cosmological terms'' in the $\beta$-function of the graviton with the
 conventional
vertex operator leading to Einstein gravity around the flat vacuum.
In case of the vertex operator (24), the OPE of 
$T_{z{\bar{z}}}^{matter}$ with $V_{grav}^{H_{-3}\otimes{H_{-3}}}$ still results in 
appearance of
the linearized Ricci tensor. However, since this operator is the element of
${H_{-3}\otimes{{\bar{H}}_{-3}}}$, and its canonical $\phi$-ghost 
picture is $(-3,-3)$ 
~{\selfsigma},
the  contributions from  
 $T_{z{\bar{z}}}^{b-c}$ and $T_{z{\bar{z}}}^{\beta-\gamma}$  
no longer cancel each other as

\eqn\lowen{(T_{z{\bar{z}}}^{b-c}+T_{z{\bar{z}}}^{\beta-\gamma})(z,{\bar{z}})
V_{grav}^{H_{-3}\otimes{H_{-3}}}(w,{\bar{w}})
\sim{{{1\over2}({1-3^2})V_{grav}^{H_{-3}\otimes{H_{-3}}}}\over{|z-w|^2}}}
leading to the cosmological term proportional to ${\sim}4G_{mn}$ 
in the $\beta$-function.
Thus the Weyl invariance condition brings  the  piece proportional to
$\sim{R^{linearized}_{mn}+4g_{mn}}$ to the $\beta$-function 
(assuming that the dilaton is switched off).
The higher order (quadratic) terms in $\beta_{mn}$ are  given by the appropriate
3-point functions.
In the next section we shall analyze these terms by computing the corresponding 
 3-point correlators on the disc.

\centerline{\bf 4. Graviton's $\beta$-function: quadratic contributions }

We start with the analysis of $<1-1-2>$ and $<3-3-2>$ correlators on the disc. 
These correlators
give rise to contributions of 
zero and second powers in momentum, particularly producing
 terms corresponding to stress tensor of ideal fluid and second order 
hydrodynamics (in this
paper we disregard the higher order contributions, 
such as those of the quartic order). 
The first order terms
stem from Weyl invariance constraints on the operators while the third order is
is produced by $<2-2-3>$ and $<2-2-1>$ disc correlators.
(in this paper, however, we do not consider the third order terms).
We start with the $<1-1-2>$ contribution. 
The spin 1 vertex operator is the element
of $H_1$, given by
\eqn\grav{\eqalign{
V_{s=1}=u_mL^m(p)=
K\circ\int{d{{z}}}e^{{\bar\phi}}
\lbrace{\lambda}\partial^2{X^a}-2\partial\lambda
\partial
{X^a}
\cr+
ip^a({1\over2}\partial^2\lambda+{1\over{q}}
\partial\varphi\partial\lambda
-{1\over2}\lambda(\partial\varphi)^2+
(1+3q^2)\lambda(3\partial\psi_b\psi^b-{1\over{2q}}
\partial^2\varphi))\rbrace{e^{ipX}}}}
To ensure the overall $\phi$-ghost number balance ($-2$ on the disc) it is 
convenient to 
take the graviton's operator unintegrated at $(-3,-3)$ picture representations
while transforming both of the integrated spin 1 operators to picture 2.
The full expression for $V_{s=1}$ at picture 2 is complicated, however we 
don't need all the terms but only those contributing to the 3-point $<2-1-1>$ 
correlator according
to ghost number selection rules. The picture 2 operator contains 
three classes of such terms
- those proportional to $e^{2\phi}$ ghost factor, those proportional
 to $be^{3\phi-\chi}$ and
those proportional to $c{e^\chi}$, so the non-vanishing ghost correlators 
are proportional
to the exponential factors ${\sim}<e^{-3\phi-3\bar\phi}(0)
ce^{\chi+\phi}(\tau_1)be^{3\phi-\chi}(\tau_2)>$
and ${\sim}<e^{-3\phi-3\bar\phi}(0)e^{2\phi}(\tau_1)e^{2\phi}(\tau_2)>$ where $\tau_1$ 
and $\tau_2$
are the locations  of the $s=1$ operators.
Straightforward evaluation of the picture-changing transformation
 of (24), however, shows that the overall
coefficient in front of the terms proportional to $be^{3\phi+\chi}$ vanishes, 
so it is only the
second ghost structure  ${\sim}<e^{-3\phi-3\bar\phi}(0)e^{2\phi}
(\tau_1)e^{2\phi}(\tau_2)>$ that is relevant to the 
correlator.
Thus we only need the part of $V_{s=1}$ at picture 2 proportional to 
$e^{2\phi}$; straightforward application
of picture-changing and homotopy transformations lead to the
 following expression for the relevant part
of $V_{s=1}$:
\eqn\grav{\eqalign{V_{s=1}(z;p)
\cr
=
u^m(p)\sum_{k=1}^5{1\over{192\times(5-k)!}}\int{d\tau}
(w-\tau)^4{e^{2\phi+ipX}}P^{(4)}_{2\phi-2\chi-\sigma}
P^{(5-k)}_{\phi-\chi}
L_m^{(k)}(\lambda,\varphi,X,\psi,\tau)}}
where $P^{(N)}_{a_1\phi+a_2\chi+a_3\chi}$ ($a_{1,2,3}$ are numbers) are the conformal
 dimension
$N$ ghost polynomials which definition and properties are 
discussed in ~{\selfp};
the space-time vectors
 $L_m^{(k)};k=1,...,5$ are the conformal dimension $k$ operators 
consisting of the matter fields,
which manifest expressions are given by:
\eqn\grav{\eqalign{
L_m^{(k)}={\lbrace}
{1\over{(k-1)!}}\partial^{(k-1)}\psi_m\lambda-
{1\over{(k-2)!}}\partial^{(k-2)}\psi_m\partial\lambda(1-\delta_1^k)-
\cr
{i\over{(k-3)!}}{\vec{p}}\partial^{(k-3)}{\vec{\psi}}F_ma(1-\delta_1^k)a
(1-\delta_2^k)
\cr
+3ip_m(1+3Q^2)\lambda(1-\delta_1^k)\lbrack
{1\over{(k-2)!}}\partial^{(k-1)}{\vec{X}}{\vec{\psi}}
-{1\over{(k-3)!}}\partial^{(k-2)}{\vec{X}}
\partial{\vec{\psi}}(1-\delta_2^k)\rbrack
\cr
-{i\over{2(k-1)!}}p_m\partial^{(k)}\varphi
+(1-\delta_1^k)\lbrack
-{1\over{(k-2)!}}\partial^{(p-1)}\varphi\partial{X_m}
+{{ip_m}\over{2Q}}
\partial^{(p-1)}\varphi\partial\varphi
\rbrack
\cr
+(1-\delta_1^k)(1-\delta_2^k)\lbrack
{1\over{2(k-3)!}}
\partial^{(k-2)}\varphi\partial^2{X_m}
\cr
+{{ip_m}\over{4}}\partial^{(p-2)}\varphi
((\partial\varphi)^2+2(1+3Q^2)(\partial{\vec{\psi}}{\vec{\psi}}
-{1\over{2Q}}\partial^2\varphi))\rbrack
\cr
+ip_m{{(1-\delta_1^k)}\over{(k-2)!}}\lbrack
{1\over{2Q}}{\lbrack}
\partial^{(k-2)}\lambda\partial\lambda-{1\over2}\partial^{(k-2)}
\lambda\lambda\partial\varphi
-{{1+3Q^2}\over{4Q(k-1)}}\partial^{(p-1)}\lambda\lambda{\rbrack}
\cr
+\delta_k^1\rbrack
-{3\over2}(1+3Q^2)p_m\lambda({\vec{p}}{\vec{\psi}})-(2+3Q^2)ip_m
\partial\varphi-2Q\partial{X_m}\rbrack
\cr
+\delta_k^2{\lbrack}
{3\over2}(1+3Q^2)p_m\lambda({\vec{p}}\partial{\vec{\psi}})+
{Q\over2}\partial^2{X_m}-
{{iQp_m}\over{4}}(\partial\varphi)^2\rbrack
{\rbrace}e^{ipX}}}
Although the expression (29) for the integrated picture 2 $V_{s=1}$
depends on an arbitrary point $z$ on the worldsheet, this dependence is
irrelevant in correlation functions since
all the $w$-derivatives of (29) are BRST-exact. For this reason, $w$ 
can be chosen
arbitrarily in the integral (29).

We are now prepared to analyze
the three-point $<2-1-1>$ amplitude on the disc.
The unintegrated $V_{s=2}$ vertex is convenient to place at the disc's origin,
 that is, at the zero point.
The calculation strategy is similar to the one described in ~{\selfswed} 
It is convenient to map a disc to a half-plane
using the conformal transformation:
\eqn\lowen{z\rightarrow{f(z)}={i\over2}{{z+i}\over{z-i}}}
and to calculate the 3-point correlator on the plane . 
The integrals over the disc boundary are 
then transformed into integrals over the real line. On the half-plane,
 it is convenient
to choose $w_1=w_2={i\over2}$ in $\tau_1$ and $\tau_2$ 
integrals for the open string vertices.
Having calculated the half-plane correlators, we shall further
 conformally map it back to
the disc and evaluate the integrals (which essentually will 
become the angular integrals).
Under the transformation (31) the left part of the $V_{s=2}$ vertex operator 
is mapped to
$z_1={i\over2}$ while the right part is mapped to $z_2=-{i\over2}$.
The ghost factors of the correlator for each term in the sum
over $k_1,k_2$ (stemming from the summation over $k$ in(29)) are given by:

\eqn\grav{\eqalign{A^{(k_1,k_2)}_{ghost}(p,k,q)
\cr
=<ce^{-3\phi}({i\over2})ce^{-3\phi}(-{i\over2})e^{2\phi}P^{(4)}_{2\phi-2\chi-\sigma}
P^{(5-k_1)}_{\phi-\chi}(\tau_1)e^{2\phi}P^{(4)}_{2\phi-2\chi-\sigma}
P^{(5-k_2)}_{\phi-\chi}(\tau_2)>
\cr
=|{i\over2}-\tau_1|^{12}|{i\over2}-\tau_2|^{12}
\times
H^{(5-k_1)}_{-3;-3;2}(\tau_1|{i\over2},-{i\over2},\tau_2)
H^{(5-k_2)}_{-3;-3;2}(\tau_2|{i\over2},-{i\over2},\tau_1)
\cr\times
{\lbrack}
H^{(4)}_{-5;-5;4}(\tau_1|{i\over2},-{i\over2},\tau_2)
H^{(4)}_{-5;-5;4}(\tau_2|{i\over2},-{i\over2},\tau_1)
\cr
+12(\tau_1-\tau_2)^2
H^{(3)}_{-5;-5;4}(\tau_1|{i\over2},-{i\over2},\tau_2)
H^{(3)}_{-5;-5;4}(\tau_2|{i\over2},-{i\over2},\tau_1)\rbrack}}
where the functions $H^{(N)}_{a_1,...a_N}(\tau|\tau_1,...\tau_N)$ 
are defined according to
\eqn\grav{\eqalign{
H^{(N)}_{a_1,...a_N}(\tau|\tau_1,...\tau_N)=
N!\sum_{N|m_1,...,m_N}^{m_1+...+m_N=N}\prod_{j=1,{m_j\neq{0}}}^N
{1\over{m_jP_{m_j}!}}\sum_{i=1}^N{{a_i}\over{(\tau_i-\tau)^{m_j}}}}}
Here $\lbrace{m_1},...{m_{N}}\rbrace;m_1<{m_2}...<{m_N}$ 
are the partitions of number $N$ of length $N$ including zeroes;
$P_{m_j}$ for $m_j\neq{0}$ are the multiplicities at which given $m_j$ 
enter the partition; and by
definition $P_0\equiv{P_{m_j=0}}\equiv{0},P_0!=1$ no matter how many zeroes 
enter the partition.
For example, the partition $10=0+0+0+0+0+1+1+2+3+3$ would read as 
$m_1=0;m_2=1,m_3=2,m_4=3$
with $P_{m_1}=P_0=0;P_{m_2}{\equiv}P_1=2;P_{m_3}{\equiv}P_2=1;P_{m_4}\equiv{P_3}=2$.
Therefore the overall $<2-1-1>$ correlator on the halfplane is given by:
\eqn\grav{\eqalign{<V_{s=2}({i\over2}, -{i\over2})V_{s=1}({i\over2})V_{s=1}
({i\over2})>=
g^{m_1m_2}(p)u^{n_1}(q_1)u^{n_2}(q_2)
\cr\times
\sum_{k_1=1}^5\sum_{k_2=1}^5{1\over{192^2\times(5-k_1)!(5-k_2)!}}
\cr\times\int_{-\infty}^\infty{d\tau_1}\int_{-\infty}^\infty{d\tau_2}
({i\over2}-\tau_1)^4({i\over2}+\tau_2)^4
\lbrack{{({i\over2}+\tau_1)({i\over2}+\tau_2)}
\over{({i\over2}-\tau_1)({i\over2}-\tau_2)}}\rbrack^{k_1k_2}
\cr\times
|{i\over2}-\tau_1|^{12}|{i\over2}-\tau_2|^{12}
\times
H^{(5-k_1)}_{-3;-3;2}(\tau_1|{i\over2},-{i\over2},\tau_2)
H^{(5-k_2)}_{-3;-3;2}(\tau_2|{i\over2},-{i\over2},\tau_1)
\cr\times
{\lbrack}
H^{(4)}_{-5;-5;4}(\tau_1|{i\over2},-{i\over2},\tau_2)
H^{(4)}_{-5;-5;4}(\tau_2|{i\over2},-{i\over2},\tau_1)
\cr
+12(\tau_1-\tau_2)^2
H^{(3)}_{-5;-5;4}(\tau_1|{i\over2},-{i\over2},\tau_2)
H^{(3)}_{-5;-5;4}(\tau_2|{i\over2},-{i\over2},\tau_1)\rbrack
\cr\times
<F_{m_1}(p;{i\over2})F_{m_2}(p;-{i\over2})L_{n_1}^{(k_1)}(q_1;\tau_1)
L_{n_2}^{(k_2)}(q_2;\tau_2)>}}
The final step is to evaluate the matter part of the correlator in (34), 
given by
$<F_{m_1}(p;{i\over2})F_{m_2}(p;-{i\over2})L_{n_1}^{(k_1)}(q_1;\tau_1)
L_{n_2}^{(k_2)}(q_2;\tau_2)>$, with the expressions for  $L_n^{(k)}$
given in (30). 
It is convenient
 to define the following functions
\eqn\grav{\eqalign{R_m^{(a)}(y|(x_1,p_1);(x_2,p_2),(x_3,p_3))
\cr
=
(-1)^a(a-1)!\lbrack{{{ip_{1m}}\over{(y-x_1)^a}}+{{ip_{2m}}\over{(y-x_2)^a}}
+{{ip_{3m}}\over{(y-x_3)^a}}}\rbrack
\cr
K_\lambda^{(a_1,a_2,a_3,a_4)}(z_1,z_2,\tau_1,\tau_2)
\cr
\equiv
<\partial^{(a_1)}\lambda(z_1)\partial^{(a_2)}\lambda(z_2)
\partial^{(a_3)}\lambda(\tau_1)
\partial^{(a_4)}\lambda(\tau_2)>\cr
=
{{(-1)^{a_1+a_3}(a_1+a_2)!(a_3+a_4)!}\over{(z_1-z_2)^{a_1+a_2+1}
(\tau_1-\tau_2)^{a_3+a_4+1}}}
\cr
+
{{(-1)^{a_1+a_2}(a_1+a_3)!(a_2+a_4)!}\over{(z_1-\tau_1)^{a_1+a_3+1}
(z_2-\tau_2)^{a_2+a_4+1}}}
\cr
+
{{(-1)^{a_1+a_2}(a_1+a_3)!(a_2+a_4)!}\over{(z_1-\tau_2)^{a_1+a_4+1}
(z_2-\tau_1)^{a_2+a_3+1}}}
\cr
S^{m_1m_2n_1n_2}_{{\lbrack}a,b,c,d{\rbrack}}(z_1,z_2,\tau_1,\tau_2)
\cr
=
{{(-1)^{a+c}\eta^{m_1m_2}\eta^{n_1n_2}}\over{(z_1-z_2)^{a+b}(\tau_1-\tau_2)^{c+d}}}
\cr
+
{{(-1)^{a+b}\eta^{m_1n_1}\eta^{m_2n_2}}\over{(z_1-\tau_1)^{a+c}(z_2-\tau_2)^{b+d}}}
\cr
+
{{(-1)^{a+b}\eta^{m_1n_2}\eta^{m_2n_1}}\over{(z_1-\tau_2)^{a+d}(z_2-\tau_1)^{b+c}}}
}}

Then the
straightforward computation gives (with $z_1={\bar{z}}_2={i\over2}$):

\eqn\grav{\eqalign{
<F_{m_1}(p;{i\over2})F_{m_2}(p;-{i\over2})L_{n_1}^{(k_1)}(q_1;\tau_1)
L_{n_2}^{(k_2)}(q_2;\tau_2)>
\cr
=\sum_{l=1}^{24}H_{m_1m_2n_1n_2}^{(l)}(z_1,z_2,\tau_1,\tau_2|p,q_1,q_2)}}

where

\eqn\grav{\eqalign{H_{m_1m_2n_1n_2}^{(1)}(z_1,z_2,\tau_1,\tau_2|p,q_1,q_2)
\cr
=
\sum_{a_1,a_2=1}^2\sum_{b_1,b_2=0}^1(\delta_{a_1}^2-2\delta_{a_1}^1)(\delta_{a_2}^2
-2\delta_{a_2}^1)
\cr\times
(\delta_{b_1}^0-\delta_{b_1}^1)(\delta_{b_2}^0-\delta_{b_2}^1)
\cr
\times{\lbrace}
{\lbrack}
{{(-1)^{a_1+a_2+b_2+k_1}\eta_{n_1n_2}}\over{(k_1-b_1-1)!(k_2-b_2-1)!
(\tau_1-\tau_2)^{p_1+p_2-b_1-b_2+1}}}
\cr\times
({(a_1+a_2-1)!\eta_{m_1m_2}\over{(z_1-z_2)^{a_1+a_2}}}
\cr+
R_{m_1}^{(a_1)}(z_1|(z_2,p);(\tau_1,q_1),(\tau_2,q_2))
R_{m_2}^{(a_2)}(z_2|(z_1,p);(\tau_1,q_1),(\tau_2,q_2)))
\cr
{\times}
({{(4-a_1-a_2)!(b_1+b_2)!}\over{(z_1-z_2)^{5-a_1-a_2}(\tau_1-\tau_2)^{b_1+b_2+1}}}
\cr-
{{(2-a_1+b_1)!(2-a_2+b_2)!}\over{(z_1-\tau_1)^{3-a_1-b_2}(z_2-\tau_2)^{-a_2+b_1+3}}}
\cr+
{{(2-a_1+b_2)!(2-a_2+b_1)!}\over{(z_1-\tau_1)^{3-a_1-b_1}(z_2-\tau_2)^{-a_2+b_2+3}}}
)\rbrack
\cr
-(1-\delta_{k_1}^1)(1-\delta_{k_2}^2)
{{iq_2^{n_1}}\over{(k_2-3)!}}
\cr{\times}{{(-1)^{k_1-b_1}(k_1+k_2-b_1-3)!K_\lambda^{(2-a_1,2-a_2,b_1,2-b_2)}
(z_1,z_2,\tau_1,\tau_2)}\over{(\tau_1-\tau_2)^{k_1+k_2-b_1-2}}}
\cr\times
({{(-1)^{a_1}(a_1+a_2-1)!\eta_{m_1m_2}}\over{(z_1-z_2)^{a_1+a_2}}}
R_{n_2}^{(b_2)}(\tau_2|(z_1,p);(z_2,p);(\tau_1,q_1))
\cr
+
{{(-1)^{a_1}(a_1+b_2-1)!\eta_{m_1n_2}}\over{(z_1-\tau_2)^{a_1+b_2}}}
R_{m_2}^{(a_2)}(\tau_1|(z_1,p);(z_2,p);(\tau_2,q_2))
\cr
+
{{(-1)^{a_2}(a_2+b_2-1)!\eta_{m_2n_2}}\over{(z_2-\tau_2)^{a_2+b_2}}}
R_{m_2}^{(a_1)}(z_1|(z_2,p);(\tau_1,q_1);(\tau_2,q_2))
)
\rbrace
}}

Next,

\eqn\grav{\eqalign{
H_{m_1m_2n_1n_2}^{(2)}(z_1,z_2,\tau_1,\tau_2|p,q_1,q_2)
\cr
=(1-\delta_{k_1}^0)
\sum_{a_1,a_2=1}^2\sum_{b_1=0}^1(\delta_{a_1}^2-2\delta_{a_1}^1)
(\delta_{a_2}^2-2\delta_{a_2}^1)
\cr\times
(\delta_{b_1}^0-\delta_{b_1}^1(1-\delta_{p_1}^1))
\cr\times
{{(-1)^{a_1+b_1+k_1+k_2}(3+9Q^2)\eta_{m_1m_2}q_{2n_1}q_{2n_2}
K_\lambda^{2-a_1;2-a_2;b_1;0}(z_1,z_2,\tau_1,\tau_2)}
\over{2(k_1-b_1-1)!(\tau_1-\tau_2)^{k_1+k_2-1-b_1}(z_1-z_2)^{a_1+a_2}}}
}}
Next,
\eqn\grav{\eqalign{
H_{m_1m_2n_1n_2}^{(3)}(z_1,z_2,\tau_1,\tau_2|p,q_1,q_2)
\cr
=(1-\delta_{k_1}^0)(1-\delta_{k_2}^0)(1-\delta_{k_2}^1)
\cr\times
\sum_{a_1,a_2=1}^2\sum_{b_1,b_2=0}^1(\delta_{a_1}^2-2\delta_{a_1}^1)
(\delta_{a_2}^2-2\delta_{a_2}^1)
\cr\times
(\delta_{b_2}^0-\delta_{b_2}^1(1-\delta_{k_2}^2))
(\delta_{b_1}^0-\delta_{b_1}^1(1-\delta_{k_1}^1))
\cr\times
{{3iq_{2n_2}(1+3Q^2)(-1)^{k_1-b_1}
K_\lambda^{2-a_1;2-a_2;b_1;0}(z_1,z_2,\tau_1,\tau_2)}\over
{(k_1-b_1-1)!(k_2-b_2-2)!(\tau_1-\tau_2)^{k_1-b_1+b_2}}}
\cr\times
{\lbrack}
{{(-1)^{a_1}\eta_{m_1m_2}
R_{n_1}^{(k_2-b_2-1)}(\tau_2|(p,z_1),(p,z_2),(q_1,\tau_1))}\over{(z_1-z_2)^{a_1+a_2}}}
\cr+
{{(-1)^{a_1}\eta_{m_1n_1}
R_{m_2}^{(a_2)}(z_2|(p,z_1),(q_1,\tau_1),(q_2,\tau_2))}\over{(z_1-z_2)^{a_1+k_2-b_2-1}}}
\cr
+
{{(-1)^{a_2}\eta_{m_2n_1}
R_{m_1}^{(a_1)}(z_1|(p,z_2),(q_1,\tau_1),(q_2,\tau_2))}\over{(z_1-z_2)^{a_2+k_2-b_2-1}}}
}}

Next,

\eqn\grav{\eqalign{
H_{m_1m_2n_1n_2}^{(4)}(z_1,z_2,\tau_1,\tau_2|p,q_1,q_2)
\cr
=(1-\delta_{k_1}^0)(1-\delta_{k_2}^0)(1-\delta_{k_1}^1)
(1-\delta_{k_2}^1)(1-\delta_{k_1}^2)
(1-\delta_{k_2}^2)
\cr\times
\sum_{a_1,a_2=1}^2\sum_{b_1,b_2=0}^1
(\delta_{a_1}^2-2\delta_{a_1}^1)(\delta_{a_2}^2-2\delta_{a_2}^1)
\cr\times
(\delta_{b_1}^2-2\delta_{b_1}^1)(\delta_{b_2}^2-2\delta_{b_2}^1)
\cr\times
{{(-1)^{k_1}(q_1q_2)K_\lambda^{2-a_1;2-a_2;2-b_1;2-b_2}(z_1,z_2,\tau_1,\tau_2)
S_{m_1m_2n_1n_2}^{{\lbrack}a_1;a_2;b_1;b_2}
(z_1,z_2,\tau_1,\tau_2)}\over{(k_1-3)!(k_2-3)!(\tau_1-\tau_2)^{k_1+k_2-5}}}}}
Next,
\eqn\grav{\eqalign{
H_{m_1m_2n_1n_2}^{(5)}(z_1,z_2,\tau_1,\tau_2|p,q_1,q_2)
\cr
=
3(1+3Q^2)q_1^nq_{2n_2}(1-\delta_{k_2}^0)
\cr\times
\sum_{a_1,a_2=1}^2\sum_{b_1=1}^2\sum_{b_2=0}^1(\delta_{a_1}^2-2\delta_{a_1}^1)
(\delta_{a_2}^2-2\delta_{a_2}^1)
\cr\times
(\delta_{b_1}^2-2\delta_{b_1}^1)
(\delta_{b_2}^0-\delta_{b_2}^1(1-\delta_{k_2}^2))
\cr\times
{{(-1)^{k_1}K_\lambda^{2-a_1;2-a_2;2-b_1;0}(z_1,z_2,\tau_1,\tau_2)
S_{m_1m_2n_1n}^{{\lbrack}a_1;a_2;b_1;k_2-b_2-1{\rbrack}}
(z_1,z_2,\tau_1,\tau_2)}\over{(k_1-3)!(k_2-2-b_2)(\tau_1-\tau_2)^{k_1+b_2-2}}}
}}
Next,
\eqn\grav{\eqalign{
H_{m_1m_2n_1n_2}^{(6)}(z_1,z_2,\tau_1,\tau_2|p,q_1,q_2)
\cr
=
-9(1+3Q^2)q_{1n_1}q_{2n_2}(1-\delta_{k_1}^0)(1-\delta_{k_2}^0)
(1-\delta_{k_1}^1)(1-\delta_{k_2}^1)
\cr\times
\sum_{a_1,a_2=1}^2\sum_{b_1,b_2=0}^1
(\delta_{a_1}^2-2\delta_{a_1}^1)(\delta_{a_2}^2-2\delta_{a_2}^1)
(\delta_{b_1}^0-\delta_{b_1}^1(1-\delta_{k_1}^2))
(\delta_{b_2}^0-\delta_{b_2}^1(1-\delta_{k_2}^2))
\cr\times
{{(-1)^{b_1}(b_1+b_2)!K_\lambda^{2-a_1;2-a_2;1;1}(z_1,z_2,\tau_1,\tau_2)
\eta^{mn}S_{m_1m_2mn}^{{\lbrack}a_1;a_2;k_1-b_1-1;k_2-b_2-1{\rbrack}}
(z_1,z_2,\tau_1,\tau_2)}\over{(k_1-b_1-2)!(k_2-b_2-2)!
(\tau_1-\tau_2)^{b_1+b_2+1}}}}}
\eqn\grav{\eqalign{
H_{m_1m_2n_1n_2}^{(7)}(z_1,z_2,\tau_1,\tau_2|p,q_1,q_2)=
-{1\over4}q_{1n_1}q_{2n_2}
\sum_{a_1,a_2=1}^2
(\delta_{a_1}^2-2\delta_{a_1}^1)(\delta_{a_2}^2-2\delta_{a_2}^1)
\cr\times
{{(-1)^{k_1}(a_1+a_2-1)!(4-a_1-a_2)!(k_1+k_2-1)!}\over
{(k_1-1)!(k_2-1)!(z_1-z_2)^5(\tau_1-\tau_2)^{p_1+p_2}}}
}}
\eqn\grav{\eqalign{
H_{m_1m_2n_1n_2}^{(8)}(z_1,z_2,\tau_1,\tau_2|p,q_1,q_2)=
{1\over2}q_{1n_1}q_{2n_2}
\cr\times
{{(-1)^{k_1}k_1(1-\delta_{k_1}^0)(1-\delta_{k_1}^1)(1-\delta_{k_2}^0)
(1-\delta_{k_2}^1)(1-\delta_{k_2}^2)}\over{(k_2-2)!(\tau_1-\tau_2)^{k_1+1}}}
\cr\times
\sum_{a_1,a_2=1}^2
(\delta_{a_1}^2-2\delta_{a_1}^1)(\delta_{a_2}^2-2\delta_{a_2}^1)
{{(-1)^{a_2}(a_1+a_2-1)!}\over{(z_1-z_2)^{a_1+a_2}}}
\cr\times
\lbrack
{{(k_2-a_2)!(2-a_1)!}\over{(z_1-\tau_2)^{k_2-a_2+1}(z_2-\tau_2)^{3-a_1}}}
-
{{(k_2-a_1)!(2-a_2)!}\over{(z_1-\tau_2)^{k_2-a_1+1}(z_2-\tau_2)^{3-a_2}}}
\rbrack}}
\eqn\grav{\eqalign{
H_{m_1m_2n_1n_2}^{(9)}(z_1,z_2,\tau_1,\tau_2|p,q_1,q_2)
\cr
=
(1-\delta_{k_1}^0)(1-\delta_{k_1}^1)(1-\delta_{k_2}^0)
(1-\delta_{k_2}^1)
\cr\times
\sum_{a_1,a_2=1}^2\sum_{b_1,b_2=0}^1
(\delta_{a_1}^2-2\delta_{a_1}^1)(\delta_{a_2}^2-2\delta_{a_2}^1)
(-\delta_{b_1}^0+{1\over2}\delta_{b_1}^1)(-\delta_{b_2}^0+{1\over2}\delta_{b_2}^1)
\cr\times
{{(-1)^{k_1+a_1+b_1}(k_1+k_2-b_1-b_2-3)!(4-a_1-a_2)!}\over{(z_1-z_2)^{5-a_1-a_2}
(\tau_1-\tau_2)^{k_1+k_2-b_1-b_2-2}}}
\times
\lbrace
S_{m_1m_2n_1n_2}^{a_1;a_21+b_1;1+b_2}
\cr
+
{{(-1)^{a_1}(a_1+b_1)!\eta_{m_1n_1}
R_{m_2}^{(a_2)}(z_2|(p,z_1);(q_1,\tau_1);(q_2,\tau_2))
R_{n_2}^{(1+b_2)}(\tau_2|(p,z_1);(p,z_2);(q_1,\tau_1))}\over{(z_1-\tau_1)^{a_1+b_1+1}}}
\cr
+
{{(-1)^{a_2}(a_2+b_1)!\eta_{m_2n_1}
R_{m_1}^{(a_1)}(z_1|(p,z_2);(q_1,\tau_1);(q_2,\tau_2))
R_{n_2}^{(1+b_2)}(\tau_2|(p,z_1);(p,z_2);(q_1,\tau_1))}\over{(z_2-\tau_1)^{a_2+b_1+1}}}
\cr
+
{{(-1)^{a_1}(a_1+b_2)!\eta_{m_1n_2}
R_{m_2}^{(a_2)}(z_2|(p,z_1);(q_1,\tau_1);(q_2,\tau_2))
R_{n_1}^{(1+b_1)}(\tau_1|(p,z_1);(p,z_2);(q_2,\tau_2))}\over{(z_1-\tau_2)^{a_1+b_2+1}}}
\cr
\rbrace}}
\eqn\grav{\eqalign{
H_{m_1m_2n_1n_2}^{(10)}(z_1,z_2,\tau_1,\tau_2|p,q_1,q_2)
\cr
=
(1-\delta_{k_1}^0)(1-\delta_{k_1}^1)(1-\delta_{k_2}^0)
(1-\delta_{k_2}^1)
\cr\times
{{iq_{2n_2}}\over{2Q(k_1-b_1-2)!(k_2-2)!}}\sum_{a_1,a_2=1}^2\sum_{b_1=0}^1
(\delta_{a_1}^2-2\delta_{a_1}^1)(\delta_{a_2}^2-2\delta_{a_2}^1)
(-\delta_{b_1}^0+{1\over2}\delta_{b_1}^1)
\cr\times
\lbrack
{{(k_1-b_1-1)!}\over{(\tau_1-\tau_2)^{k_1-b_1}}}\times
{{(k_2-a_1)!(2-a_2)!}\over{(z_1-\tau_2)^{k_2-a_1+1}(z_2-\tau_2)^{3-a_2}}}
-
{{(k_2-a_2)!(2-a_1)!}\over{(z_2-\tau_2)^{k_2-a_2+1}(z_1-\tau_2)^{3-a_1}}}
\rbrack
\cr\times
\lbrack
{{(-1)^{a_2+1}(a_1+b_1)!\eta_{m_1n_1}
R_{m_2}^{(a_2)}(z_2|(p,z_1);(q_1,\tau_1);(q_2,\tau_2))}\over
{(z_1-\tau_1)^{a_1+b_1+1}}}
\cr
+
{{(-1)^{a_1+1}(a_1+b_1)!\eta_{m_2n_1}
R_{m_1}^{(a_1)}(z_1|(p,z_2);(q_1,\tau_1);(q_2,\tau_2))}\over
{(z_2-\tau_1)^{a_2+b_1+1}}}}}
\eqn\grav{\eqalign{
H_{m_1m_2n_1n_2}^{(11)}(z_1,z_2,\tau_1,\tau_2|p,q_1,q_2)
\cr
=
-(1-\delta_{k_1}^0)(1-\delta_{k_1}^1)(1-\delta_{k_2}^0)
(1-\delta_{k_2}^1)\times
{{iq_{2n_2}(2+3Q^2)}\over{(k_1-b_1-2)!(k_2-2)!}}
\cr\times
\sum_{a_1,a_2=1}^2\sum_{b_1=0}^1
(\delta_{a_1}^2-2\delta_{a_1}^1)(\delta_{a_2}^2-2\delta_{a_2}^1)
(-\delta_{b_1}^0+{1\over2}\delta_{b_1}^1)
\cr\times
\lbrack
{{(-1)^{k_1-b_1}(k_1-b_1-1)!}\over{(\tau_1-\tau_2)^{k_1-b_1}}}\times
{{(-1)^{a_1}(4-a_1-a_2)!}\over{(z_1-z_2)^{5-a_1-a_2}}}
\rbrack
\cr\times
\lbrack
{{(-1)^{a_1}(a_1+b_1)!\eta_{m_1n_1}
R_{m_2}^{(a_2)}(z_2|(p,z_1);(q_1,\tau_1);(q_2,\tau_2))}\over
{(z_1-\tau_1)^{a_1+b_1+1}}}
\cr
+
{{(-1)^{a_2}(a_1+b_1)!\eta_{m_2n_1}
R_{m_1}^{(a_1)}(z_1|(p,z_2);(q_1,\tau_1);(q_2,\tau_2))}\over
{(z_2-\tau_1)^{a_2+b_1+1}}}
\rbrack
}}
\eqn\grav{\eqalign{
H_{m_1m_2n_1n_2}^{(12)}(z_1,z_2,\tau_1,\tau_2|p,q_1,q_2)=
{{ip_{m_2}}\over2}\eta_{n_1n_2}(1-\delta_{k_1}^0)(1-\delta_{k_2}^0)
\cr\times
\sum_{a_1}^2\sum_{b_1,b_2=0}^1
{{(\delta_{a_1}^2-2\delta_{a_1}^1)
(\delta_{b_1}^0-\delta_{b_1}^1(1-\delta_{p_1}^1))
(\delta_{b_2}^0-\delta_{b_2}^1(1-\delta_{p_2}^1))}\over{(k_1-b_1-1)!(k_2-b_2-1)!}}
\cr\times
(-1)^{k_1-b_1}(k_1+k_2-b_1-b_2-2)!{{K_\lambda^{2-a_1;2;b_1;b_2}
R_{m_1}^{(a_1)}(z_1|(p,z_2);(q_1,\tau_1);
(q_2,\tau_2))}\over{(\tau_1-\tau_2)^{k_1+k_2-b_1-b_2-1}}}}}
\eqn\grav{\eqalign{
H_{m_1m_2n_1n_2}^{(13)}(z_1,z_2,\tau_1,\tau_2|p,q_1,q_2)
\cr
=
{1\over2}q_{2n_1}p_{m_2}\eta_{m_1n_2}
(1-\delta_{k_1}^0)(1-\delta_{k_1}^1)(1-\delta_{k_2}^0)
(1-\delta_{k_2}^1)(1-\delta_{k_2}^2)
\cr\times
\sum_{a_1}^2\sum_{b_1,b_2=0}^1
{{(\delta_{a_1}^2-2\delta_{a_1}^1)(\delta_{a_2}^2-2\delta_{a_2}^1)
(\delta_{b_1}^0-\delta_{b_1}^1(1-\delta_{p_1}^1))
(\delta_{b_2}^0-\delta_{b_2}^1(1-\delta_{p_2}^1))}\over{(k_1-b_1-1)!(k_2-3)!}}
\cr\times
(-1)^{a_1+b_1+k_1}(a_1+a_2-1)!
{{K_\lambda^{2-a_1;2;b_1;2-a_2}(z_1,z_2,\tau_1,\tau_2)
(k_1+k_2-b_1-3)!}\over{(z_1-\tau_2)^{a_1+a_2}(\tau_1-\tau_2)^{k_1+k_2-b_1-2}}}}}
\eqn\grav{\eqalign{
H_{m_1m_2n_1n_2}^{(14)}(z_1,z_2,\tau_1,\tau_2|p,q_1,q_2)
\cr
=
-{3\over2}(1+3Q^2)q_{2n_2}p_{m_2}\eta_{m_1n_1}
(1-\delta_{k_1}^0)(1-\delta_{k_1}^1)(1-\delta_{k_2}^0)
(1-\delta_{k_2}^1)
\cr\times
\sum_{a_1}^2\sum_{b_1,b_2=0}^1
{(\delta_{a_1}^2-2\delta_{a_1}^1)
(\delta_{b_1}^0-\delta_{b_1}^1(1-\delta_{p_1}^1))
(\delta_{b_2}^0-\delta_{b_2}^1(1-\delta_{p_2}^1))}
\cr\times
(-1)^{a_1+b_1+k_1}(a_1+k_2-b_2-2)!
{{K_\lambda^{2-a_1;2;b_1;0}(z_1,z_2,\tau_1,\tau_2)
(k_1+b_2-b_1)!}\over{(z_1-\tau_2)^{a_1+k_2-b_2-1}(\tau_1-\tau_2)^{k_1+b_2-b_1+1}}}
}}
\eqn\grav{\eqalign{
H_{m_1m_2n_1n_2}^{(15)}(z_1,z_2,\tau_1,\tau_2|p,q_1,q_2)
\cr
=
p_{m_2}q_{1n_1}\eta_{m_1n_2}
{{(1-\delta_{k_1}^0)(1-\delta_{k_1}^1)(1-\delta_{k_2}^0)
(1-\delta_{k_2}^1)}\over{(k_1-1)!}}
\cr\times
\sum_{a_1}^2\sum_{b_1=0}^1
{{(\delta_{a_1}^2-2\delta_{a_1}^1)
(-\delta_{b_1}^0+{1\over2}\delta_{b_1}^1(1-\delta_{k_1}^1))}\over{(k_2-b_2-2)!}}
\cr\times
{{(a_1+b_2)!}\over{(z_1-\tau_2)^{a_1+b_2+1}}}
\times
{\lbrack}
{1\over2}
{{(2-a_1)!k_1!(k_2-1-b_2)!}\over{(z_1-z_2)^{3-a_1}(z_2-\tau_1)^{k_1+1}
(z_2-\tau_2)^{k_2-b_2}}}
\cr
+{1\over4}(-1)^{k_1+1}
{{(4-a_1)!(k_1+k_2-2-b_2)!}\over{(z_1-z_2)^{5-a_1}(\tau_1-\tau_2)^{k_1+k_2-b_2-1}}}
\rbrack
}}
\eqn\grav{\eqalign{
H_{m_1m_2n_1n_2}^{(16)}(z_1,z_2,\tau_1,\tau_2|p,q_1,q_2)=
{{p_{m_1}q_{1n_1}\eta_{m_2n_1}}\over{(k_1-1)!}}
\cr\times
\sum_{a_1}^2
(\delta_{a_1}^2-2\delta_{a_1}^1)(-1)^{a_1+1}
{{(3-a_1)!k_1!}\over{(z_1-z_2)^{4-a_1}(z_2-\tau_1)^{k_1+1}}}
\cr\times
\lbrack
{1\over2}\delta_{k_2}^2{{(a_1+1)!}\over{(z_1-\tau_2)^{a_1+2}}}-2\delta_{k_2}^1
{{a_1!}\over{(z_1-\tau_2)^{a_1+1}}}
\rbrack
}}
\eqn\grav{\eqalign{
H_{m_1m_2n_1n_2}^{(17)}(z_1,z_2,\tau_1,\tau_2|p,q_1,q_2)
\cr
=
(1-\delta_{k_1}^0)(1-\delta_{k_2}^0)(1-\delta_{k_1}^1)(1-\delta_{k_2}^1)
(-ip_{m_2})
\cr\times
\sum_{a_1=1}^2
\sum_{b_1,b_2=0}^1
{(\delta_{a_1}^2-2\delta_{a_1}^1)
(-\delta_{b_1}^0+{1\over2}\delta_{b_1}^1(1-\delta_{k_1}^2))
(-\delta_{b_2}^0+{1\over2}\delta_{b_2}^1(1-\delta_{k_2}^2))}
\cr\times
{{(k_1-1-b_1)(k_1-1-b_2)}\over{(z_2-\tau_1)^{k_1-b_1}(z_1-\tau_2)^{k_2-b_2}}}
\cr\times
{{\eta_{m_1n_1}(a_1+b_1)!R_{n_2}^{(1+b_2)}(\tau_2|(p,z_1);(p,z_2);(q_1,\tau_1))}\over
{(z_1-\tau_1)^{a_1+b_1+1}}}
\cr
+
{{\eta_{m_1n_2}(a_1+b_2)!R_{n_1}^{(1+b_1)}(\tau_1|(p,z_1);(p,z_2);(q_2,\tau_2))}\over
{(z_2-\tau_2)^{a_1+b_2+1}}}
}}
\eqn\grav{\eqalign{
H_{m_1m_2n_1n_2}^{(18)}(z_1,z_2,\tau_1,\tau_2|p,q_1,q_2)
\cr
=-p_{m_2}q_{2n_2}\eta_{m_1n_1}
(1-\delta_{k_1}^0)(1-\delta_{k_2}^0)(1-\delta_{k_1}^1)(1-\delta_{k_2}^1)
\cr\times
\sum_{a_1,a_2=1}^2
\sum_{b_1,b_2=0}^1
(\delta_{a_1}^2-2\delta_{a_1}^1)
(-\delta_{b_1}^0+{1\over2}\delta_{b_1}^1(1-\delta_{k_1}^2))
\cr\times
{{(-1)^{a_1}(a_1+b_1)!}\over{(k_1-2-b_1)!(z_1-\tau_1)^{a_1+b_1+1}}}
\cr\times
{{{1\over{{Q^2}}}(-\delta_{a_2}^1+{{1+3Q^2}\over2}\delta_{a_2}^2)
({1\over2}\delta_{b_2}^0+{1\over8}\delta_{b_2}^1(1-\delta_{k_2}^2))}
}
\cr\times\lbrace
{{(-1)^{a_1+a_2+k_1+b_1}(4-a_1-a_2)!}\over{(k_1-b_1-2)!(z_1-z_2)^{5-a_1-a_2}}}
\cr\times
\lbrack
{{(k_2+a_2-b_2-2)!(k_1-b_1+b_2-1)!}\over{(z_2-\tau_2)^{k_2+a_2-b_2-1}
(\tau_1-\tau_2)^{k_1-b_1+b_2}}}
\cr
+
{{(a_2+b_2)!(k_1+k_2-b_1-b_2-3)!}\over{(z_2-\tau_2)^{a_2+b_2+1}
(\tau_1-\tau_2)^{k_1+k_2-b_1-b_2-2}}}
\rbrack
\cr
+
{{(-1)^{a_1+k_1+b_1}(1-\delta_{k_2}^2)}\over{8(k_2-3)!(z_1-z_2)^{3-a_1}}}
\cr\times\lbrack
{{2(k_1+k_2-b_1-4)!}\over{(z_2-\tau_2)^{4}(\tau_1-\tau_2)^{k_1+k_2-b_1-3}}}
+
{{2(k_2-2)!(k_1-b_1-1)!}\over{(z_2-\tau_2)^{k_2+1}(\tau_1-\tau_2)^{k_1-b_1}}}
\rbrack
\cr
+{{3(1+3Q^2)(-1)^{a_1+b_1+k_1}(1-\delta_{k_2}^2)(k_1+k_2-b_1-4)!}
\over{(k_2-3)!(z_1-z_2)^{3-a_1}(z_2-\tau_2)^4(\tau_1-\tau_2)^{k_1+k_2-b_1-3}}}
\rbrace
}}
\eqn\grav{\eqalign{
H_{m_1m_2n_1n_2}^{(19)}(z_1,z_2,\tau_1,\tau_2|p,q_1,q_2)
\cr=-p_{m_2}q_{2n_2}\eta_{m_1n_1}
(1-\delta_{k_1}^0)(1-\delta_{k_2}^0)(1-\delta_{k_1}^1)
\cr\times
\sum_{a_1,a_2=1}^2
\sum_{b_1,b_2=0}^1
{{(\delta_{a_1}^2-2\delta_{a_1}^1)
(-\delta_{b_1}^0+{1\over2}\delta_{b_1}^1(1-\delta_{k_1}^2))}
\over{(k_1-2-b_1)!}}
\cr\times
(-{{1+3Q^2}\over{4Q}}\delta_{b_2}^0+{1\over{2Q}}\delta_{b_2}^1
(1-\delta_{k_2}^2))
({1\over{Q}}\delta_{a_2}^1-{{1+3Q^2}\over{2Q}}\delta_{a_2}^2)
\cr\times\lbrace
{{(a_1+b_1)!(a_2+k_1-b_1-2)!}\over{(z_1-\tau_1)^{a_1+b_1+1}
(z_2-\tau_1)^{a_2+k_1-b_1-1}}}
\cr\times
{\lbrack}
{{(1-a_1+k_2-b_2)!(2-a_2+b_2)!}\over{(z_1-\tau_2)^{2-a_1+k_2-b_2}
(z_2-\tau_2)^{3+b_2-a_2}}}
\cr-
{{(1-a_2+k_2-b_2)!(2-a_1+b_2)!}\over{(z_1-\tau_2)^{2-a_2+k_2-b_2}
(z_2-\tau_2)^{3+b_2-a_1}}}
\rbrack
\cr
+{{(-1)^{a_1}(1-\delta_{k_2}^1)(k_1-b_1-1)!}
\over{(z_2-\tau_1)^{k_1-b_1}(z_2-\tau_2)^{2}}}
\cr\times\lbrack
{{(k_2-a_1)!}\over{(z_1-\tau_2)^{k_2-a_1+1}(z_2-\tau_2)}}
-{{(k_2-2)!}\over{(z_1-\tau_2)^{3-a_1}(z_2-\tau_2)^{k_2-1}}}
\rbrack
\rbrace
}}
\eqn\grav{\eqalign{
H_{m_1m_2n_1n_2}^{(20)}(z_1,z_2,\tau_1,\tau_2|p,q_1,q_2)
\cr
=
(1-\delta_{k_1}^0)(1-\delta_{k_2}^0)(1-\delta_{k_1}^1)\delta_{k_2}^1
\cr\times
\sum_{a_1=1}^2
\sum_{b_1=0}^1
{{(\delta_{a_1}^2-2\delta_{a_1}^1)
(-\delta_{b_1}^0+{1\over2}\delta_{b_1}^1(1-\delta_{k_1}^2))}
\over{(k_1-2-b_1)!}}
\cr\times
\lbrace
(2+3Q^2)p_{m_2}q_{2n_2}\eta_{m_1n_1}\times
{{(-1)^{a_1}(a_1+b_1)!}\over{2(z_1-\tau_1)^{a_1+b_1+1}}}
\cr\times
\lbrack
{{(-1)^{a_1+b_1+k_1}(4-a_1)!(k_1-b_1-1)!}\over{(z_1-z_2)^{5-a_1}
(\tau_1-\tau_2)^{k_1-b_1}}}
\cr
-2{{(-1)^{a_1+1}(k_1-b_1-1)!}
\over{(z_1-z_2)^{3-a_1}(z_2-\tau_1)^{k_1-b_1}(z_2-\tau_2)^2}}
\rbrack
\cr
+\sum_{a_2=1}^2(-2\delta_{a_2}^1+(1+3Q^2)\delta_{a_2^2})
(ip_{m_2})
\cr\times
{{(-1)^{a_1+a_2+1}(4-a_1-a_2)!(k_1-b_1+a_2-2)!}\over
{(z_1-z_2)^{5-a_1-a_2}(z_2-\tau_1)^{k_1-b_1+a_2-1}}}
\cr\times\lbrack
{{(-1)^{a_1}(a_1+b_1)!\eta_{m_1n_1}R_{n_2}^{(1)}(\tau_2|(p,z_1);(p,z_2);(q_1,\tau_1))}
\over{(z_1-\tau_1)^{a_1+b_1+1}}}
\cr
+
{{(-1)^{a_1}a_1!\eta_{m_1n_2}R_{n_1}^{(1+b_1)}(\tau_1|(p,z_1);(p,z_2);(q_2,\tau_2))}
\over{(z_1-\tau_2)^{a_1+1}}}
\cr
+
{{(-1)^{1+b_1}(1+b_1)!\eta_{n_1n_2}R_{m_1}^{(a_1)}(z_1|(p,z_2);(q_1,\tau_1);(q_2\tau_2))}
\over{(\tau_1-\tau_2)^{b_1+2}}}
\rbrack
\rbrace
}}
\eqn\grav{\eqalign{
H_{m_1m_2n_1n_2}^{(21)}(z_1,z_2,\tau_1,\tau_2|p,q_1,q_2)
\cr=
(1-\delta_{k_1}^0)(1-\delta_{k_2}^0)(1-\delta_{k_1}^1)\delta_{k_2}^2
\cr\times
\sum_{a_1,a_2=1}^2
\sum_{b_1=0}^1
{{(\delta_{a_1}^2-2\delta_{a_1}^1)
(-\delta_{b_1}^0+{1\over2}\delta_{b_1}^1(1-\delta_{k_1}^2))}
\over{(k_1-2-b_1)!}}
\cr{\times}
{{(\delta_{a_2}^1-{1\over2}(1+3Q^2)\delta_{a_2^2})
(ip_{m_2})(-1)^{a_1+1}(4-a_1-a_2)!}\over{(z_1-z_2)^{5-a_1-a_2}}}
\cr\times\lbrace
{{(-1)^{a_2}(a_2+k_1-b_1-2)!}\over{(z_2-\tau_1)^{a_2+k_1-b_1-1}}}
\lbrack
{{(-1)^{a_1}(a_1+b_1)!\eta_{m_1n_1}R_{n_2}^{(2)}
(\tau_2|(p,z_1);(p,z_2);(q_1,\tau_1))}\over
{2(z_1-\tau_1)^{a_2+k_1-b_1-1}}}
\cr
+
{{(-1)^{a_1}(a_1+1)!\eta_{m_1n_2}R_{n_1}^{(1+b_1)}(\tau_1|(p,z_1);(p,z_2);
(q_2,\tau_2))}\over
{2(z_1-\tau_2)^{a_1+2}}}
\cr
+
{{(-1)^{1+a_1}(b_1+2)!\eta_{m_1n_2}R_{m_1}^{(a_1)}(z_1|(p,z_2);(q_1,\tau_1);(q_2\tau_2))}
\over{2(\tau_1-\tau_2)^{b_1+3}}}\rbrack
\cr
+
{{iq_{2n_2}\eta_{m_1n_1}(-1)^{a_1+a_2+b_1+k_1}(a_1+b_1)!a_2!(k_1-b_1-1)!}\over
{2(z_1-\tau_1)^{a_1+b_1+1}(z_2-\tau_2)^{a_2+1}(\tau_1-\tau_2)^{k_1-b_1}}}
\rbrace
}}
\eqn\grav{\eqalign{
H_{m_1m_2n_1n_2}^{(22)}(z_1,z_2,\tau_1,\tau_2|p,q_1,q_2)=
(1-\delta_{k_1}^0)(1-\delta_{k_1}^1)
\cr\times
\sum_{a_1=1}^2
\sum_{b_1=0}^1
(\delta_{a_1}^2-2\delta_{a_1}^1)
({{\delta_{b_1}^0}\over{2Q}}+{{(1+3Q^2)(1-\delta_{k_1}^2)\delta_{b_1}^1}\over{4Q}})
\cr
\times{{p_{m_2}q_{1n_1}\eta_{m_1n_2}(-1)^{a_1}(k_1-1-b_1)!(1+b_1)!}
\over{(z_1-z_2)^{3-a_1}(z_2-\tau_1)^{k_1}}}
\times\lbrack
{{Q\delta_{k_2}^2(a_1+1)!}\over{2(z_1-\tau_2)^{a_1+2}}}
-{{2Q\delta_{k_2}^1{a_1}!}\over{(z_1-\tau_2)^{a_1+1}}}
\rbrack
}}
\eqn\grav{\eqalign{
H_{m_1m_2n_1n_2}^{(23)}(z_1,z_2,\tau_1,\tau_2|p,q_1,q_2)=
-(1-\delta_{k_1}^0)(1-\delta_{k_1}^1)(1-\delta_{k_1}^2)
\cr\times{{3(1+3Q^2)}\over{2(k_1-3)!}}\times
{{\eta_{m_1n_2}p_{m_2}q_{1n_1}}\over{(z_2-\tau_1)^4}}
\cr\times
\sum_{a_1=1}^2
(\delta_{a_1}^2-2\delta_{a_1}^1)
{{(-1)^{a_1+1}}\over{(z_1-z_2)^{3-a_1}}}
\lbrack
{{(-1)^{a_1}Q\delta_{k_2}^2(a_1+1)!}\over{2(z_1-\tau_2)^{a_1+2}}}
-{{2Q\delta_{k_2}^1{a_1}!}\over{(z_1-\tau_2)^{a_1+1}}}
\rbrack
}}

\eqn\grav{\eqalign{
H_{m_1m_2n_1n_2}^{(24)}(z_1,z_2,\tau_1,\tau_2|p,q_1,q_2)
\cr=
-p_{m_2}q_{1n_1}\eta_{m_1n_2}
\times
\sum_{a_1=1}^2
\sum_{b_1=0}^1
(\delta_{a_1}^2-2\delta_{a_1}^1)
\times\lbrace
{{(-{{1+3Q^2}\over{4Q}}\delta_{b_1}^0+{1\over{2Q}}(1-\delta_{k_1}^1)
\delta_{b_1}^1)}\over
{2(k_1-b_1-1)!}}
\cr\times
{\lbrack}
-{{Q\delta_{k_2}^2(a_1+1)!}\over{2(z_1-\tau_2)^{a_1+2}}}
+(-1)^{a_1}{{2Q\delta_{k_2}^1{a_1}!}\over{(z_1-\tau_2)^{a_1+1}}}
\rbrack
\cr
\times
\lbrack
{{(k_1-a_1-b_1+1)!(b_1+2)!}\over{(z_1-\tau_1)^{k_1-a_1-b_1+2}(z_2-\tau_1)^{3+b_1}}}
-{{(2-a_1+b_1)!(k_1+1-b_1)!}\over{(z_1-\tau_1)^{3-a_1+b_1}(z_2-\tau_1)^{p_1+2-b_1}}}
\rbrack
\cr
+
\sum_{a_2=0}^2
\lbrack
({{\delta_{a_2}^1}\over{Q}}-{{1+3Q^2}\over{2Q}}\delta_{a_2}^2)
\times{{1-\delta_{p_1}^1}\over{2(k_1-2)!}}
\times{{a_2!}\over{(\tau_1-\tau_2)^{a_2+1}}}
\rbrack
\cr\times
{\lbrack}
-{{Q\delta_{k_2}^2(a_1+1)!}\over{2(z_1-\tau_2)^{a_1+2}}}
+(-1)^{a_1}{{2Q\delta_{k_2}^1{a_1}!}\over{(z_1-\tau_2)^{a_1+1}}}
\rbrack\times
{\lbrack}
-{{Q\delta_{k_2}^2(a_1+1)!}\over{2(z_1-\tau_2)^{a_1+2}}}
+(-1)^{a_1}{{2Q\delta_{k_2}^1{a_1}!}\over{(z_1-\tau_2)^{a_1+1}}}
\rbrack
\cr
\times\lbrack
{{(k_1-a_1)!}\over{(z_1-\tau_1)^{k_1-a_1+1}(z_2-\tau_1)^{3-a_2}}}
-
{{(k_1-a_2)!}\over{(z_1-\tau_1)^{3-a_1}(z_2-\tau_1)^{k_1+1-a_2}}}
\rbrack
\rbrace}}

\eqn\grav{\eqalign{
H_{m_1m_2n_1n_2}^{(25)}(z_1,z_2,\tau_1,\tau_2|p,q_1,q_2)=
\delta_{k_1}^1\delta_{k_2}^1\sum_{a_1,a_2=1}^2(\delta_{a_1}^2-2\delta_{a_1}^1)
(2\delta_{a_2}^2-(1+3Q^2)\delta_{a_2}^1)
\cr\times
\lbrack
-(1+3Q^2)p_{m_2}q_{1n_1}\eta_{m_1n_2}
{{(-1)^{a_2+1}a_1!a_2!(4-a_1-a_2)!}\over{(z_1-z_2)^{5-a_1-a_2}
(z_1-\tau_2)^{a_1+1}(z_2-\tau_1)^{a_2+1}}}
\cr
+2Q^2(-1)^{a_1+1}(ip_{m_2})\times{{(4-a_1)!}\over{(z_1-z_2)^{5-a_1}}}
\times
({{(-1)^{a_1}a_1!\eta_{m_1n_1}
R_{n_2}^{(1)}(\tau_2|(p,z_1);(p,z_2);(q_1,\tau_1))}\over{(z_1-\tau_1)^{a_1+1}}}
\cr
+
{{(-1)^{a_1}a_1!\eta_{m_1n_2}R_{n_1}^{(1)}(\tau_1|(p,z_1);(p,z_2);(q_2,\tau_2))}\over
{(z_1-\tau_2)^{a_1+1}}}
-
{{\eta_{n_1n_2}R_{m_1}^{(a_1)}(z_1|(p,z_2);(q_1,\tau_1);(q_2\tau_2))}\over
{(\tau_1-\tau_2)^2}}
)
\rbrack
}}

\eqn\grav{\eqalign{
H_{m_1m_2n_1n_2}^{(26)}(z_1,z_2,\tau_1,\tau_2|p,q_1,q_2)=
\delta_{k_1}^2\delta_{k_2}^2\sum_{a_1=1}^2(\delta_{a_1}^2-2\delta_{a_1}^1)
\cr\times
(ip_{m_2})\times{{(4-a_1)!}\over{(z_1-z_2)^{5-a_1}}}
\times
({{(-1)^{a_1}(a_1+1)!\eta_{m_1n_1}
R_{n_2}^{(2)}(\tau_2|(p,z_1);(p,z_2);(q_1,\tau_1))}\over{(z_1-\tau_1)^{a_1+2}}}
\cr+
{{(-1)^{a_1}a_1!\eta_{m_1n_2}R_{n_1}^{(2)}(\tau_1|(p,z_1);(p,z_2);(q_2,\tau_2))}\over
{(z_1-\tau_2)^{a_1+2}}}
+
{{6\eta_{n_1n_2}R_{m_1}^{(a_1)}(z_1|(p,z_2);(q_1,\tau_1);(q_2\tau_2))}\over
{(\tau_1-\tau_2)^4}}
)
\cr
+{{p_{m_2}q_{2n_2}\eta_{m_1n_1}(a_1+1)!}\over{(z_1-\tau_1)^{a_1+2}(z_1-z_2)^{3-a_1}
(z_2-\tau_2)^4}}
}}
\eqn\grav{\eqalign{
H_{m_1m_2n_1n_2}^{(27)}(z_1,z_2,\tau_1,\tau_2|p,q_1,q_2)=
\delta_{k_1}^1\delta_{k_2}^2\sum_{a_1=1}^2(\delta_{a_1}^2-2\delta_{a_1}^1)
\cr\times\lbrace
{{ip_{m_2}Q^2(-1)^{a_1+1}(4-a_1)!}\over{2(z_1-z_2)^{5-a_1}}}
\times
({{(-1)^{a_1}a_1!\eta_{m_1n_1}
R_{n_2}^{(2)}(\tau_2|(p,z_1);(p,z_2);(q_1,\tau_1))}\over{(z_1-\tau_1)^{a_1+1}}}
\cr
+
{{(-1)^{a_1}(a_1+1)!\eta_{m_1n_2}R_{n_1}^{(1)}
(\tau_1|(p,z_1);(p,z_2);(q_2,\tau_2))}\over
{(z_1-\tau_2)^{a_1+2}}}
\cr-
{{\eta_{n_1n_2}R_{m_1}^{(a_1)}(z_1|(p,z_2);(q_1,\tau_1);(q_2\tau_2))}\over
{(\tau_1-\tau_2)^2}})
\cr
-{{Qp_{m_2}q_{2n_2}\eta_{m_1n_1}{a_1}!}\over{2(z_1-z_2)^{3-a_1}(z_1-\tau_1)^{a_1+1}
(z_2-\tau_2)^{4}}}
\cr
+Qp_{m_2}q_{1n_1}\eta_{m_1n_2}\sum_{a_2=1}^2({{\delta_{a_2}^1}\over{Q}}
-{{1+3Q^2}\over{2Q}}\delta_{a_2}^2)
{{(-1)^{a_2+1}(a_1+1)!a_2!(4-a_1-a_2)!}\over{(z_1-z_2)^{5-a_1-a_2}(z_1-\tau_2)^{a_1+2}
(z_2-\tau_1)^{a_2+1}}}
\rbrace
}}
This gives the $<2-1-1>$ correlator, contributing to the graviton's 
$\beta$-function.
Next, consider the contributions from spin 3 excitations, 
that stem from the $<2-3-3>$
correlator.
The spin 3 vertex operators are given by
\eqn\grav{\eqalign{V_{s=3}^{(-3)}=H_{mab}(q)ce^{-3\phi}\psi^{m}
\partial{X^{a}}\partial{X^{b}}e^{iqX}(z)}}
at negative unintegrated cohomology $H_{-3}$ representation
and 
\eqn\grav{\eqalign{V_{s=3}^{(+1)}=H_{mab}(q)K\circ\oint{dz}e^{\phi}\psi^{m}
\partial{X^{a}}\partial{X^{b}}e^{iqX}(z)}}
at positive $H_{1}$ cohomology representation.
The on-shell conditions on the  spin 3 field $H_{mab}$ are given by
\eqn\grav{\eqalign{q^aH_{mab}=0}}
\eqn\grav{\eqalign{
\eta^{ab}H_{mab}=0}}
and
\eqn\grav{\eqalign{
\eta^{ma}H_{mab}=0}}

As was noted above, instead of considering $H_{mab}$ as fundamental excitations,
we are looking for polynomial combinations of $u$ coupling to the 
vertex operators (64), (65) and satisfying the on-shell conditions 
(66)-(68) to ensure
the BRST properties of the operators.
The only suitable combination satisfying (66)-(68) is given by
\eqn\grav{\eqalign{H_{mab}(p)=
\int{d^4k}\int{d^4q}u_m(k+q)u_a(k-p)u_b(q-p)
\cr
+{1\over2}\delta_{ab}u_m(p)-{1\over2}(\delta_{ma}u_b(p)+
\delta_{mb}u_a(p))}}
In order to satisfy (66)-(68)
 $u_a$ must furthermore satisfy $u_au^a=-1$ with zero vorticity condition
$p_{\lbrack{a}}u_{b\rbrack}(p)=0$ and incompressibility $p_au^a(p)=0$.
Note that  the $u^2=-1$ constraint can also be obtained
from the vanishing on the $\beta$-function  for the spin 1 operator
(29) which, in the leading order, can be computed to give
$\beta^a_{u_a}\sim{u_b}(g^{ab}+u^au^b)$.
As the $\beta$-function is the object that must be calculated off-shell,
in the calculations below we shall keep
the terms, that are both non-transverse and have nonzero vorticities, as they
only vanish in the on-shell limit.

As in the $<2-1-1>$-computation,
it is convenient to take the graviton's operator unintegrated
at canonical $(-3,-3)$-picture, locating it at the disc's origin
(accordingly, at $z={i\over2}$ on the half-plane).
As for spin 3 operators, located at the boundary of the disc
(accordingly, on the real line after the transformation
to the half-plane)
they both should therefore
be taken integrated at picture $+2$.
Instead transforming the operator (65) to picture 2 
by picture-changing transform, it is more convenient
to consider the operator 
\eqn\grav{\eqalign{V^{2|1}=2\omega_n^{a_1a_2|b}V_{a_1a_2|b}^n(p)\cr
V^n_{a_1a_2|b}(p)=K\circ\oint{e^{2\phi}}
(-2\partial\psi^m\psi_b\partial{X_{(a_1}}\partial^2{X_{a_2)}}
\cr
-2 \partial\psi^m\partial\psi_b\partial{X_{a_1}}\partial{X_{a_2}}
+
\psi^m\partial^2\psi_b\partial{X_{a_1}}\partial{X_{a_2}})e^{ipX}}}
with $V^{2|1}$ being a vertex operator for the spin 3
$extra$ field $\omega^{2|1}$ in Vasiliev's frame-like
formalism ~{\selfp}. This extra field is related to
the dynamical metric-like field $\omega^{2|0}\equiv{H_{nab}}$ of (65)
up to BRST-exact terms
through the cohomology constraint ~{\selfp} given by
\eqn\grav{\eqalign{
\omega_n^{ab|c}(p)=2p^cH^{ab}_n(p)-p^aH^{bc}_n(p)-p^bH^{ac}_m(p)}}
In addition, it is straightforward to check  that the Weyl 
invariance of $V^{2|1}$ also
requires 
\eqn\lowen{\omega_c^{ab|c}=0}
which can also be seen directly from the primary field constraint on
 $V^{2|1}$ at the dual
$-4$ picture.
In particular, this implies that the graviton's $\beta$-function 
can be shifted according to
\eqn\grav{\eqalign{\beta^{mn}\rightarrow\beta^{mn}+const\times\omega_c^{mn|c}}}
since such a shift corresponds to the same on-shell limit and, in this limit,
doesn't violate the conformal invariance on the worldsheet.
Next, given (69) and (71), the vanishing  $\omega_c^{ab|c}$ condition 
(72) leads to
\eqn\grav{\eqalign{-p_{(a}u_{b)}+\int{d^4k}\int{d^4q}u^m(k+q)p_mu_a(k-p)u_b(q-p)=0}}
or, in the position space,
\eqn\grav{\eqalign{\partial_{(a}u_{b)}+u_{(a}({\vec{u}}{\vec{\partial}})u_{b)}=0}}

Note that, with the $u^2=-1$ constraint the left-hand side of (75) can 
be cast as  the traceless tensor, transverse with respect to $u^a$:

$$\omega_c^{ab|c}\sim{\Pi^{ac}}\Pi^{bd}\partial_{(c}u_{d)}-
{1\over3}{\Pi^{ab}}(\partial_cu^c)$$
which is nothing but the first-derivative dissipative term in the
 hydrodynamical stress tensor.

The straightforward calculation of the $<2-3-3>$-correlator then gives
\eqn\grav{\eqalign{
<V_{s=2}({i\over2}, -{i\over2})V_{s=3}({i\over2})V_{s=3}({i\over2})>
\cr
=g^{m_1m_2}(p)
(2q_1^{c_1}H^{a_1b_1n_1}(q_1)-q_1^{a_1}H^{b_1c_1n_1}(q_1)
-q_1^{b_1}H^{a_1c_1n_1}(q_1))
\cr\times
(2q_2^{c_2}H^{a_1b_1n_1}(q_2)-q_2^{a_2}H^{b_2c_2n_2}(q_2)
-q_2^{b_2}H^{a_2c_2n_2}(q_2))
\cr\times
\int_{-\infty}^\infty{d\tau_1}\int_{-\infty}^\infty{d\tau_2}
({i\over2}-\tau_1)^{2-q_1q_2}({i\over2}+\tau_2)^{2-q_1q_2}
({i\over2}-\tau_2)^{6-q_1q_2}
\cr\times
({i\over2}+\tau_1)^{6-q_1q_2}
(\tau_1-\tau_2)^{q_1q_2-4}
\times
\lbrack
H^{(4)}_{-5;-5;4}(\tau_1|z_1,z_2,\tau_2)H^{(4)}_{-5;-5;4}(\tau_2|z_1,z_2,\tau_1)
\cr
+{{12}\over{(\tau_1-\tau_2)^2}}
H^{(3)}_{-5;-5;4}(\tau_1|z_1,z_2,\tau_2)H^{(3)}_{-5;-5;4}(\tau_2|z_1,z_2,\tau_1)
\rbrack
\cr\times
 \sum_{\alpha_1,\alpha_2=1}^2(\delta_{\alpha_1}^2-2\delta_{\alpha_1^1})
(\delta_{\alpha_2}^2-2\delta_{\alpha_2^1})
(-4\delta_1^{\beta_1}\delta_0^{\gamma_1}\delta_1^{\rho_1}\delta_2^{\lambda_1}
-4
\delta_1^{\beta_1}\delta_1^{\gamma_1}\delta_1^{\rho_1}\delta_1^{\lambda_1}
+2\delta_0^{\beta_1}\delta_2^{\gamma_1}\delta_1^{\rho_1}\delta_1^{\lambda_1}
)
\cr\times
(-4\delta_1^{\beta_2}\delta_0^{\gamma_2}\delta_1^{\rho_2}\delta_2^{\lambda_2}
-4
\delta_1^{\beta_2}\delta_1^{\gamma_2}\delta_1^{\rho_2}\delta_1^{\lambda_2}
+2\delta_0^{\beta_2}\delta_2^{\gamma_2}\delta_1^{\rho_2}\delta_1^{\lambda_2}
)
\cr\times
{{(-1)^{\alpha_2+\beta_1+\gamma_1+\lambda_1}(4-\alpha_1-\alpha_2)}
\over{(z_1-z_2)^{5-\alpha_1-\alpha_2}}}
\times
{{\eta_{n_1n_2}\eta_{c_1c_2}(\beta_1+\beta_2)!(\gamma_1+\gamma_2)!
-\eta_{n_1c_2}\eta_{n_2c_1}}
\over{(\tau_1-\tau_2)^{\beta_1+\beta_2+\gamma_1+\gamma_2+2}}}
\cr\times\lbrack
{{\eta_{m_1a_1}\eta_{m_2a_2}\eta_{b_1b_2}(\alpha_1+\rho_1-1)!
(\alpha_2+\rho_2-1)!(\lambda_1+\lambda_2-1)!}
\over{(z_1-\tau_1)^{\alpha_1+\rho_1}
(z_2-\tau_2)^{\alpha_2+\rho_2}(\tau_1-\tau_2)^{\lambda_1+\lambda_2}}}
\cr+
{{\eta_{m_1a_2}\eta_{m_2a_1}\eta_{b_1b_2}(\alpha_1+\rho_2-1)!
(\alpha_2+\rho_1-1)!(\lambda_1+\lambda_2-1)!}
\over{(z_1-\tau_2)^{\alpha_2+\rho_1}
(z_2-\tau_1)^{\alpha_1+\rho_2}(\tau_1-\tau_2)^{\lambda_1+\lambda_2}}}
\rbrack}}
This concludes the computation of the integrand of the
$2-3-3$ correlator contributing to the graviton's $\beta$-function.
The next step is to perform the integrations 
of the lengthy expressions (34)-(63) and (76) in $\tau_1$ and $\tau_2$. 
All the integrals entering the
$<2-1-1>$ and $<2-3-3>$ amplitudes
 (34)-(63), (76) have the form:
\eqn\grav{\eqalign{I(\alpha_1,\alpha_2;\beta_1,\beta_2;\gamma;L)
\cr=
(z_1-z_2)^{L}\int_{-\infty}^\infty{d\tau_1}\int_{-\infty}^{\tau_1}{d\tau_2}
(\tau_1-z_1)^{\alpha_1}
(\tau_1-z_2)^{\alpha_2}(\tau_2-z_1)^{\beta_1}(\tau_2-z_2)^{\beta_2}
(\tau_1-\tau_2)^{\gamma}}}
with the powers in the integrands given by
\eqn\grav{\eqalign{
\alpha_{1,2}=-2q_1q_2+M_{1,2}\cr
\beta_{1,2}=-2q_1q_2+N_{1,2}\cr
\gamma=k_1k_2+P}}
where $L,M_1,M_2,N_1,N_2,P$ being various combinations of the integer numbers
 following from the manifest expressions (34)-(63),(76), so the overall
 $<2-1-1>$ and $<2-3-3>$ amplitudes are given by the appropriate summations
\eqn\lowen{<2-s-s>|_{s=1,3}\sim\sum_{L,M_1,M_2,N_1,N_2,P}I(a_1,a_2;b_1,b_2;c;d)}
The contributions of these sums to the $AdS$ graviton's $\beta$-function
in the limit $\alpha^\prime\rightarrow{0}$ are then determined by the 
coefficients
in front of the simple pole $\sim{(q_1q_2)^{-1}}$ produced by the 
integrations.
Although the integral (77) looks complicated  one of hypergeometric type,
things simplify drastically if we use the overall conformal invariance
of the amplitudes (34),(76) allowing us to map the half-plane
expressions back to the disc. The integrals over
$\tau_1$ and $\tau_2$ are then conformally mapped
to the double angular integral over $\varphi_1,\varphi_2$
with the angular variable $0\leq\varphi\leq{2\pi}$ parametrizing
the  boundary of the disc.
With the conformal transformation (31), it is easy to check that
the relation between $\varphi$ and $\tau$ is 
\eqn\lowen{\tau={1\over2}tan({{\varphi}\over2}+{\pi\over4})}
The resulting angular integrals turn out to be remarkably
simpler. Namely, simple computation  using
the conformal transformations of (31), (80) 
and 
changing the angular variable according to
${{\varphi}\over2}+{\pi\over4}\rightarrow\varphi$
gives

\eqn\grav{\eqalign{I(\alpha_1,\alpha_2;\beta_1,\beta_2;\gamma;L)=
(-1)^{L\over2}2^{-\alpha_1-\alpha_2-\beta_1-\beta_2-\gamma}
\cr\times
{\lbrack}F(\alpha_1+\alpha_2|\beta_1+\beta_2|\gamma)
+F(\alpha_1+\alpha_2+2|\beta_1+\beta_2|\gamma)
\cr+
F(\alpha_1+\alpha_2|\beta_1+\beta_2+2|\gamma)
+
F(\alpha_1+\alpha_2|\beta_1+\beta_2|\gamma)}}

where

\eqn\grav{\eqalign{
F(\alpha|\beta|\gamma)=\int_0^\pi{d\varphi_1}
\int_0^{\varphi_1}{d\varphi_2}tan^\alpha\varphi_1
tan^\beta\varphi_2(tan\varphi_1-tan\varphi_2)^\gamma}}

Integrating one obtains
\eqn\grav{\eqalign{F(\alpha|\beta|\gamma)
=i\pi{\lbrack}{{(\delta_{-\gamma-2}^{\alpha+\beta}+\delta_{-\gamma-4}^{\alpha+\beta})
\Gamma(\beta+1)\Gamma(\gamma+1)}\over{\Gamma(\beta+\gamma+2)}}
+
{{(\delta_{-\gamma-4}^{\alpha+\beta}+\delta_{-\gamma-6}^{\alpha+\beta})
\Gamma(\beta+3)\Gamma(\gamma+1)}\over{\Gamma(\beta+\gamma+4)}}\rbrack}}
This is precisely the pole structure we are looking for.
Using Mathematica, it is now straightforward to simplify the 
integrands of (34)-(63), (76), to
substitute the appropriate values of
$I(\alpha_1,\alpha_2;\beta_1,\beta_2;\gamma;L)$ for each of the integrals
using and to compute the coefficient in front of
the pole $(k_1k_2)^{-1}$ in the field theory limit
$\alpha^\prime\rightarrow{0}$. The final result is that
the contributions of spin 1 and spin 3 excitations to the 
$\beta$-function of the graviton are given by
\eqn\grav{\eqalign{\beta^{mn}_{<2-1-1>}+\beta^{mn}_{<2-3-3>}
=\Lambda{{dg^{mn}(p)}\over{d\Lambda}}
\cr
=
32\int{d^4q}u^m(q-p)u^n(q+p)
-{1\over2}\sum_{j=1}^{10}T_j^{mn}}}
where
\eqn\grav{\eqalign{
T_1^{mn}=
{{3}}
\lbrace
\delta_a^{b_1}\delta_c^{b_2}\delta_m^{b_3}
-
\delta_c^{b_1}\delta_a^{b_2}\delta_m^{b_3}
-
\delta_a^{b_1}\delta_m^{b_2}\delta_c^{b_3}
\cr+
\delta_c^{b_1}\delta_m^{b_2}\delta_a^{b_3}
-
\delta_m^{b_1}\delta_c^{b_2}\delta_a^{b_3}
+
\delta_m^{b_1}\delta_a^{b_2}\delta_c^{b_3}
\rbrace\cr
\times\lbrace
\int{d^4k}\int{d^4q_1}\int{d^4q_2}
u^{a}(k-p)(q_1+q_2)^{(c}u^{n)}(q_1+q_2)
\cr\times
u_{b_1}(q_1-k-p)(q_2-k-p)_{b_2}u_{b_3}(q_2-k-p)
\cr
+
\int{d^4k_1}\int{d^4k_2}\int{d^4k_3}\int{d^4q_1}\int{d^4q_2}
(q_1+q_2)_du^a(k_1+k_2)u^d(k_2-p)
\cr\times
u^{(c}(k_3-k_2+p)u^{n)}(q_1+q_2)u_{b_1}(q_1-k_3-k_2+p)
\cr\times
(q_2-k_3-k_2+p)_{b_2}
u_{b_3}(q_2-k_3-k_2+p)
\rbrace}}
\eqn\grav{\eqalign{T_2^{mn}=
64\eta_{st}\int{d^4k}{\lbrace}\omega_c^{ms|c}(k-p)\omega_d^{nt|d}(k+p)
\cr
-{1\over3}\eta^{mn}\omega^{ps|c}_c(k-p)\omega^{ps|d}_d(k+p)\rbrace
\cr
+({{20}\over3}-12Q^2-{{8}\over{Q^2}})\eta^{mn}\int{d^4k}
\lbrack
(k-p)_a(k+p)^au_b(k-p)u^b(k+p)
\cr+
(k-p)_a(k+p)_bu^b(q-p)u^a(k-p)\rbrack
\cr
-{{64}\over3}
\int{d^4k}\int{d^4q_1}\int{d^4q_2}
u^m(k-p)u^n(q_1+q_2)
\cr\times
\lbrack
(q_2-k-p)_a(q_1-k-p)_bu^a(q_1-k-p)u^b(q_2-k-p)
\cr
+
(q_2-k-p)_a(q_1-k-p)^au_b(q_1-k-p)u^b(q_2-k-p)\rbrack
}}
\eqn\grav{\eqalign{
T_3^{mn}={32}\int{d^4k}(k-p)_au^a(k-p)\omega^{mn|c}_c(k+p)
}}
\eqn\grav{\eqalign{T_4^{mn}=
{{96}}\int{d^4k}\int{d^4q_1}\int{d^4q_2}
(k-p)_a(q_1+q_2)_b
\cr\times
u^m(k-p)u^n(q_1+q_2)u^a(q_1-k-p)u^b(q_2-k-p)
\cr
-
{{32}}\eta^{mn}
\int{d^4k}\int{d^4q_1}\int{d^4q_2}
(k-p)_a(q_1+q_2)_b
\cr\times
u_p(k-p)u^p(q_1+q_2)u^a(q_1-k-p)u^b(q_2-k-p)
\cr
-
{16}\eta^{mn}
\int{d^4k_1}\int{d^4k_2}\int{d^4k_3}\int{d^4q_1}\int{d^4q_2}
u^m(k_1+k_2)u^n(k_2-p)
\cr\times
(k_3-k_2+p)_a(q_1+q_2)_bu_p(k_3-k_2+p)u^p(q_1+q_2)
\cr\times
u^a(q_1-k_3-k_2+p)u^b(q_2-k_3-k_2+p)
}}
\eqn\grav{\eqalign{T_5^{mn}=
4(3Q^2-1-{{2}\over{Q^2}})
\int{d^4k}u^p(p-k)(p+k)_p({3\over2}(p+k)_mu_n(p+k)
\cr
+{3\over2}(p+k)^nu^m(p+k)-\eta^{mn}(p+k)^au_a(p+k))
}}
\eqn\grav{\eqalign{T_6^{mn}=
12
\int{d^4k}\int{d^4q_1}\int{d^4q_2}\lbrace
(q_2-k+p)_c
{u^n}(k+p)u_b(q_1+q_2)u^c(q_1-k+p)
\cr
(u^{b}(q_2-k+p)(q_2-k+p)^{m}+u^{m}(q_2-k+p)(q_2-k+p)^{b})
\rbrace
+perm{\lbrace}m\leftrightarrow{n}\rbrace}}

\eqn\grav{\eqalign{T_7^{mn}=
-16
\int{d^4k}\int{d^4q_1}\int{d^4q_2}
(u^{m}(k+p)u^{n}(q_1+q_2)+u^{n}(k+p)u^{m}(q_1+q_2))
\cr\times
u^c(q_1-k+p)u^a(q_2-k+p)
(q_2-k+p)_a(q_2-k+p)_c}}
\eqn\grav{\eqalign{T_8^{mn}=
-16
\eta^{mn}
\int{d^4k}\int{d^4q_1}\int{d^4q_2}
u^a(k+p)u^b(q_1+q_2)
\cr\times
u^c(q_1-k+p)u_b(q_2-k+p)
(q_2-k+p)_a(q_2-k+p)_c}}
\eqn\grav{\eqalign{T_9^{mn}=
48
\int{d^4k_1}\int{d^4k_2}
\int{d^4k_3}\int{d^4q_1}\int{d^4q_2}
\cr\lbrace
u^{(m}(q_1+q_2+k_1)u^{n)}(q_1+q_2-k_1)
u^a(p-q_1-k_2)u^b(p-q_1+k_2)
\cr\times
u^p(q_2-p-k_3)u_b(q_2-p+k_3)
(q_2-p+k_3)_a(q_2-p+k_3)_p\rbrace}}
\eqn\grav{\eqalign{T_{10}^{mn}=
16\int{d^4k_1}\int{d^4k_2}\int{d^4k_3}\int{d^4q_1}\int{d^4q_2}
\cr\lbrace
u^{(m}(q_1+q_2+k_1)u^{n)}(p-q_1-k_2)u^a(q_1+q_2-k_1)
u^b(p-q_1+q_2)\cr\times
u^p(q_2-p-k_3)u_b(q_2-p+k_3)
(q_2-p+k_3)_a(q_2-p+k_3)_p\rbrace}}

As it is clear from (84)-(94), the overall  result for
the $\beta$-function of the graviton, polarized along
the $d=4$ boundary, generally depends on the value $Q$ of the Liouville
background charge (which in turn can be expressed in terms of the central charge

$c_{Liouv}=1+3Q^2$). Therefore in general,
the trace of the $\beta$-function is nonzero.
Transforming (85)-(94) to the position space and using the $u^2=-1$ condition
it is straightforward to check that the overall trace of (84)-(94) vanishes
 for $Q={\sqrt{2}}$  which precisely is the case for $d+1=5$, where the
trace of spin $1$ contributions
 is cancelled by that of spin $3$.
In this case the answer  has a natural interpretation in terms
of 
holographic fluid.
This concludes
the computation of the graviton's $\beta$ function in $AdS$ string sigma-model,
up to terms quadratic in momentum.
Combining (84)-(94) and (27) one finds that
 vanishing of the beta-function (84) leads to low-energy
equations of motion in space-time, equivalent to equations gravity with the 
matter,
described by the stress-energy tensor of four-dimensional conformal fluid.
To see the relevance of this matter stress tensor to holographic 
hydrodynamics, one has to shift
 $\beta^{mn}$ according to
\eqn\grav{\eqalign{\beta^{mn}\rightarrow
{\tilde{\beta}}^{mn}=\beta^{mn}+16i\omega^{mn|c}_c}}
with $\omega_p^{mn|p}$ given by (69), (71).
As was explained above, such a shift  does not change the on-shell 
limit of the theory
due to the Weyl invariance constraint (72) on spin 3 vertex operator.
The resulting stress-energy tensor for the matter then simply describes
the conformally invariant second order hydrodynamics
at the temperature $T=\pi^{-1}$ with 
 5 extra transport coefficients in the second order. 
The relative values of the transport coefficients are
become remarkably close to those obtained in the $AdS_5$ gravity computations 
~{\mino}, 
with less then 10 percent discrepancy, albeit
at a specific temperature  in string theory calculations performed in this work.
Note that in conformal second order hydrodynamics
both the stress-energy tensor and the temperature transform covariantly under 
the $4d$ Weyl rescalings: $g_{mn}\rightarrow{e^\rho}g_{mn}$ according to
$T^{mn}\rightarrow{e^{-3\rho}}T^{mn}$ and $T\rightarrow{e^{-{\rho\over2}}}T$
so the temperature can always be fixed by appropriate Weyl transformation.
In other words, the holographic second order hydrodynamics appears in
 a particular gauge which, in a sense, is not surprising,
as it is generally the case in string theory calculations.
In the next concluding section we shall discuss the implications of the 
main result
(84) and particularly outline the calculations that still need to be done.

\centerline{\bf 5. Conclusions and Discussion}

In case of $d=4,Q_{d+1=5}={\sqrt{2}}$ the two-derivative piece
of the matter stress tensor in the graviton's $\beta$-function becomes traceless
and can be interpreted in terms of two-derivative corrections to conformal
hydrodynamics in $d=4$.
Transforming to the position space,
it is straightforward to relate the
contributions to the graviton's $\beta$-function
to corresponding terms in the gradient expansion
in conformal hydrodynamics.
 The contributions related to the Weyl invariance constraints on the graviton's
operator combined with contributions from
the $<2-1-1>$-correlator of order zero in momentum result in the ideal 
conformal fluid terms
in the $\beta$-function, proportional to $g^{mn}+4u^mu^n$. Shifting the
 $\beta$-function
 by the trace of $\omega^{2|1}$ spin 3 extra field, $\sim{\omega^{mn|p}_p}$,
which vanishing on-shell follows from the Weyl invariance constraints on the
spin 3 operator (72) leads to the leading order dissipative term, containing one
derivative due to the ghost cohomology/zero torsion constraint (71) relating
extra fields to the dynamical field in Vasiliev's formalism.
Finally, the contributions (85)-(94) given by $T^{mn}_i(i=1,...,10)$ are 
quadratic in momentum
and stem from the  $<2-3-3>$ correlator combined with the appropriate terms
from the $<2-1-1>$ correlator. These  terms describe the two-derivative 
dissipative corrections in the second order hydrodynamics ~{\rom, \mino, \mint}.
The spin $3$ contribution is crucial to ensure the vanishing trace of 
the matter tensor.
Note that, at least in the approximation considered in this paper (up to 
second order)
there are no contributions from the mixed $<2-1-3>$ correlator,
 as all the relevant terms
in this correlator are cubic in $\lambda$ and vanishing for this reason.
Transforming to the position space, 
it is straightforward to identify $T^{mn}_i$ with 
the corresponding two-derivative structures in the second order hydrodynamics,
related to $5$ new transport coefficients for the conformal fluid, appearing
in the second order. Namely, 
\eqn\grav{\eqalign{
T_1^{mn}\sim\epsilon^{bb_1b_2b_3}\epsilon^{abc(m}\rho^{n)}_cu_au_{b_1}\partial_{b_2}u_{b_3}
\cr
T_2^{mn}{\sim}3\rho^{ma}\rho^n_{a}-\eta^{mn}\rho^{ab}\rho_{ab}
\cr
T_3^{mn}\sim\rho^{mn}\partial_au^a
\cr
T_4^{mn}\sim{3}({\vec{u}}{\vec{\partial}})u^m({\vec{u}}{\vec{\partial}})u^n
-(\eta^{mn}+u^mu^n)({\vec{u}}{\vec{\partial}})u_a({\vec{u}}{\vec{\partial}})u^a
\cr
\sum_{i=5}^{10}T_i^{mn}\sim
(3\Pi^{ma}\Pi^{nb}-\Pi^{mn}\Pi^{ab})({\vec{u}}{\vec{\partial}})
(\partial_au_b+\partial_bu_a)
}}

with $\rho^{ab}\sim\omega^{ab|c}_c$

These structures are all well-known to appear in the second order of the 
gradient
expansion of the conformal fluid.
They correspond to $T_{2a},T_{2b},T_{2c}$, $T_{2d}$ and $T_{2e}$ terms, 
considered in ~{\mino}.

The correlators considered in this paper, as well as those related to graviton
interactions with operators of higher spin values  
will also contribute the higher derivative
contributions (with three and more derivatives) that were not 
addressed in this work.
At this stage, many more higher spin correlators should enter the game,  
possibly
 including those  with mixed symmetries and 
those coming from closed string sector.
As in the two-derivative case, however, 
the conformal symmetry significantly reduces
the number of terms and new transport coefficients at  higher orders. 
It is not clear at present if higher order corrections to the gradient expansion
in conformal hydrodynamics can be described in terms of
 contributions from two-row
Vasiliev's frame-like fields or more mixed symmetry 
degrees of freedom are needed
The latter almost certainly produce the structures that are present 
in the third and higher order
hydrodynamics but violate the $4d$ conformal symmetry, however the question is
whether the contributions from the two-row fields are sufficient to 
describe the conformal limit.
To answer these questions we need to have  better understanding of the 
general expansion structure of 
higher order hydrodynamics. 
Our main conjecture, based on the leading order results of this paper, suggests
that , in general, the gradient expansion in conformal hydrodynamics in $d=4$ is
controlled by the higher spin correlators in string theory and, in the leading 
$\alpha^\prime$ order, the derivative structure of the gradient expansion must 
be holographically
related to that of higher spin vertices and to the structure constants
of higher spin algebra in $AdS_5$, with the orders of the expansion roughly 
corresponding
to the total spin value carried by the HS vertices.
It would be particularly interesting to explore  the relation of the
 gradient expansion
at higher orders to well-known structures of the cubic and quartic vertices 
for higher spins
~{\vcubic, \boulanger, \bbd, \metsaev, \taronnao, \taronnas, \sagnottinew, 
\mirian,
\mirianf, \euo, \eut}
which presumably should exist in the limit of  $\alpha^\prime\rightarrow{0}$.
If the higher spin interpretation of the gradient expansion
in hydrodynamics, investigated in this paper in the string theory context, 
is still correct
at higher orders, the higher spin  algebra in $d=5$ would provide a 
powerful tool allowing
to control the transport coefficients in higher order hydrodynamics.
Another important problem to investigate is the role of $\alpha^\prime$ 
corrections 
in this expansion and their holographic interpretation. 
This may lead to new nontrivial
and intriguing symmetries relating the expansion structures and 
transport coefficients
at different orders and understanding these symmetries in terms of higher 
spin quantization.
The work on these and other issues is currently in progress and we hope to
be able to present our results soon in future works.

\centerline{\bf 6. Acknowledgements }

It is a pleasure to thank Bum-Hoon Lee, Shiraz Minwalla, 
Hermann Nicolai and Misha Vasiliev
for useful discussions and comments.

One of us (D.P.) would like to express his gratitude to Hermann Nicolai 
for kind hospitality and to
Albert Einstein Institute (AEI) in Potsdam where part of this work has been 
completed.
D.P. also would like to thank the organizers of 
the 2nd Solvay Workshop on Higher Spin Gauge Theories
at Solvay Institute at University of Brussels 
and the participants of this Workshop
for stimulating discussions.

This work was supported by the National 
Research Foundation of Korea(NRF) grant funded 
by the Korea government(MEST) through the Center for 
Quantum Spacetime(CQUeST) of 
Sogang University with grant number 2005-0049409.
S.L. and D.P. also acknowledge the support of the NRF grant number 2012-004581.

\listrefs

\end